\documentclass[aps,prl,twocolumn,reprint,superscriptaddress,nobibnotes]{revtex4-2}
\usepackage{graphicx}
\usepackage{epstopdf}
\usepackage{amssymb, amsmath}
\usepackage[usenames]{color}
\usepackage{bm}
\usepackage{ulem}
\usepackage{multirow}
\usepackage{threeparttable}
\usepackage{booktabs}
\usepackage{dcolumn}
\usepackage{float}
\usepackage[colorlinks=true, linkcolor=blue,anchorcolor=blue, citecolor=blue,urlcolor=blue]{hyperref}
\begin{document}

\title{Stacking-dependent Electronic and Topological Properties in van der Waals Antiferromagnet MnBi$_2$Te$_4$ Films}

\author{Jiaheng Li}
\affiliation{Beijing National Laboratory for Condensed Matter Physics and Institute of physics, Chinese academy of sciences, Beijing 100190, China}
\affiliation{University of Chinese academy of sciences, Beijing 100049, China}

\author{Quansheng Wu}
\email{quansheng.wu@iphy.ac.cn}
\affiliation{Beijing National Laboratory for Condensed Matter Physics and Institute of physics, Chinese academy of sciences, Beijing 100190, China}
\affiliation{University of Chinese academy of sciences, Beijing 100049, China}

\author{Hongming Weng}
\email{hmweng@iphy.ac.cn}
\affiliation{Beijing National Laboratory for Condensed Matter Physics and Institute of physics, Chinese academy of sciences, Beijing 100190, China}
\affiliation{University of Chinese academy of sciences, Beijing 100049, China}
\affiliation{Songshan Lake Materials Laboratory, Dongguan, Guangdong 523808, China}

\begin{abstract}
Combining first-principles calculations and tight-binding Hamiltonians, we study the stack-dependent behaviour of electronic and topological properties of layered antiferromagnet MnBi$_2$Te$_4$. Lateral shift of top septuple-layer greatly modify electronic properties, and even induce topological phase transition between quantum anomalous Hall (QAH) insulators with $C=1$ and trivial magnetic insulators with $C=0$. The local energy minimum of ``incorrect" stacking order exhibits thickness-dependent topology opposite to the usual stacking order, which is attribute to relatively weakened interlayer Te-Te interaction in ``incorrect" stacking configuration. Our effective model analysis provides a comprehensive understanding of the underlying mechanisms involved, and we also propose two optical setups that can effectively differentiate between different stacking configurations. Our findings underscores the nuanced and profound influence that interlayer sliding in magnetic topological materials can have on the macroscopic quantum states, opening new avenues for the design and engineering of topological quantum materials.

\end{abstract}

\maketitle

The role of stacking order in vdW layered materials is pivotal and has been empirically substantiated across diverse domains, such as sliding ferroelectricity, low-dimensional magnetism and moire superlattices  \cite{wu2021sliding, stern2021interfacial, sivadas2018stacking, chen2019direct, andrei2020graphene}. The extensively studied twisted bilayer graphene showcase the emergence of flat bands around the magic angle of approximately 1.1 degrees \cite{bistritzer2011moire}. These flat bands arise from the interaction between electronic states originating from different local stacking configurations in real space, offering a fertile ground for exploring novel quantum states and phenomena, including moire physics \cite{cao2018correlated, lu2019superconductors}, superconductivity \cite{cao2018unconventional, oh2021evidence}, and quantum anomalous Hall (QAH) effects \cite{serlin2020intrinsic}. Another typical example is bilayer CrI$_3$ film, in which interlayer sliding can induce distinct interlayer magnetic interaction from ferromagnetic to antiferromagnetic states \cite{sivadas2018stacking, li2019pressure, song2019switching, jiang2019stacking}.

Meanwhile, the recent theoretical prediction \cite{li2019intrinsic, otrokov2019unique, zhang2019topological} and experimental fabrication of MnBi$_2$Te$_4$ (MBT) \cite{lee2013crystal, gong2019experimental, otrokov2019prediction, yan2019crystal, cui2019transport, li2020antiferromagnetic}, the long-sought antiferromagnetic topological insulator \cite{mong2010antiferromagnetic}, have sparked intensive research interests \cite{li2023progress, wang2023topological, hu2023recent}, such as axion electrodynamics \cite{sekine2021axion, zhao2021routes}, and layer Hall effect \cite{gao2021layer, chen2022layer}. As a typical van der Waals (vdW) layered material, bulk MBT is characterized by its ABC-stacked Te-Bi-Te-Mn-Te-Bi-Te septuple-layer (SL) building blocks. The antiferromagnetic interlayer interaction between adjecent ferromagnetic SLs culminate in A-type antiferromagnetism. The interplay between topology and intrinsic magnetism in MBT provide a ideal material platform for layer-dependent topological properties, with odd SLs exhibiting quantum anomalous Hall (QAH) insulating behaviors, and even SLs manifesting axion insulating characteristics \cite{li2019intrinsic, otrokov2019unique, deng2020quantum, liu2020robust}. Although stacking-dependent topological phases have been recognized in MBT \cite{ren2022quantum, zhu2022high, cao2023switchable, ahn2023stacking} and chemically similar materials, including Bi$_2$Te$_3$ \cite{kou2018tunable} and GeBi$_2$Te$_4$\cite{peng2020stacking}, the exploration of various stacking patterns in MBT multilayer films remains significantly underexplored, and a unified model Hamiltonian to describe distinct stacking orders is still absent.

\begin{figure}[htbp]
	\includegraphics[width=\linewidth]{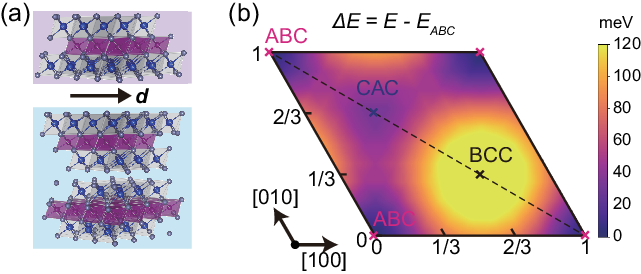}
	\caption{Atomic structure and potential energy map of three septuple layers (SL) MnBi$_2$Te$_4$ (MBT). (\textbf{a}) Crystal structure of 3-SL MBT, with Mn, Bi and Te atoms denoted as purple, blue and gray balls, respectively. Lateral shift of the bottom SLs relative to the top SL is denoted by the vector $\mathbf{d}$. (\textbf{b}) Stacking energy of 3-SL MBT as a function of lateral shift vector $\mathbf{d}$, compared to the ABC-stacking MBT.}
	\label{fig1}
\end{figure}

In this work, we theoretically predicted that the manipulation of stacking orders in MBT thin films can significantly alter their electronic and topological characteristics. Although all stacking orders can maintain interlayer antiferromagnetism, there emerge topological phase transitions between $C=1$ and $C=0$ topological phases only through lateral shift of the topmost SL. The ``incorrect" stacking order unexpectedly exhibits quantum anomalous Hall (QAH) insulators with $C=1$ in even SLs with compensated AFM phases and trivial magnetic insulators with $C=0$ in odd SLs with uncompensated AFM phases. In addition, a tight-binding model is developed to offer a comprehensive and detailed description of electronic and topological properties exhibited by various stacking configurations, and provide a valuable tool in the exploration of other relevant properties of MBT multi-layer films.

\begin{figure*}[htbp]
	\includegraphics[width=0.9\linewidth]{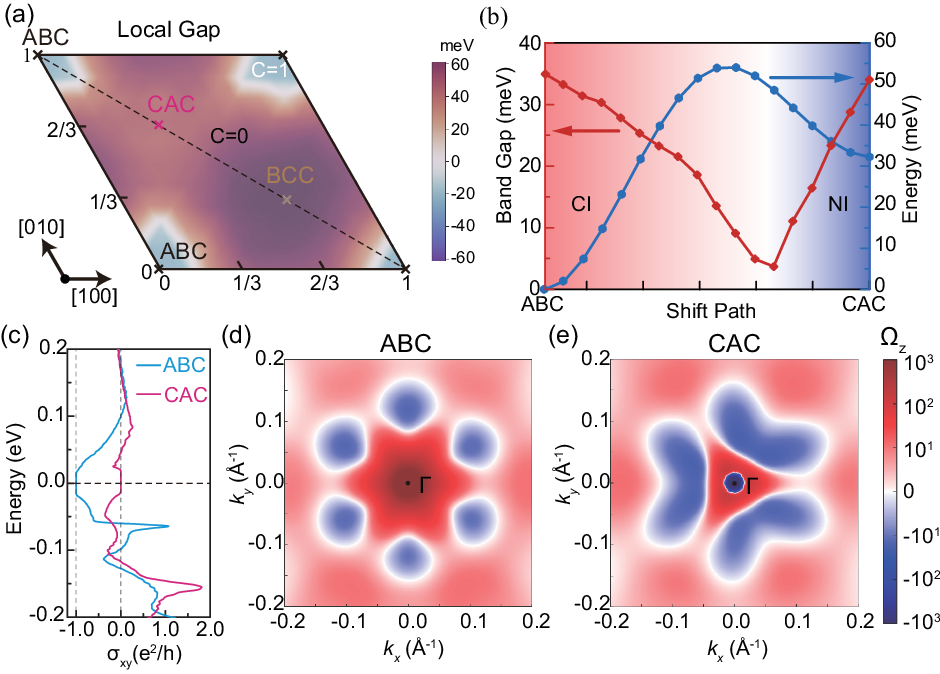}
	\caption{Electronic and topological properties of 3-SL MBT thin films. (\textbf{a}) Calculated band gap as a function of lateral shift. The positive (negative) sign of band gaps indicates quantum anomalous Hall insulators ($C=1$) and trivial magnetic insulators ($C=0$). (\textbf{b}) The evolution of band gaps and stacking energy between ABC- and CAC-stacking order along the high-symmetry [1-10]. The grey shaded area indicates the topological transition point. Anomalous Hall conductance $\sigma_{xy}$ as a function of Fermi energy (\textbf{c}) and the distribution of Berry curvature in the momentum space (\textbf{d}, \textbf{e}) in the ABC- and CAC-stacking order, respectively.}
	\label{fig2}
\end{figure*}

Taking a representative example, the crystal structure of 3-SL MBT is plotted in Fig. \ref{fig1}(a), where the lateral shift of the bottom SLs relative to the top SL is denoted by the vector $\mathbf{d}$. The potential energy surface of 3-SL MBT with different stacking orders in Fig. \ref{fig1}(b) is evaluated with reference to the ground magnetic state. The ABC stacking order is at the global minimum, and the CAC-stacking is energetically comparable and also located at local minimum. Here, the ABC-stacking is the usual stacking order in MBT bulk forms and thin films, and the CAC-stacking refers to the ``incorrect" stacking order that can be obtained by shifting the remaining bottom SLs by [1/3, -1/3] relative to the top SL (Fig. \ref{fig1}(a)). This above notation also applies to other MBT films, such as CABC- and BCAC-stacking in 4-SL MBT. Our further phonon dispersions \cite{SM} also validate that CAC-stacking MBT is dynamically stable, implying that 3-SL MBT thin films can be trapped in the CAC stacking order. Contrary to ABC- and CAC-stacking, the BBC-stacking is energetically unfavorable, and will spontaneously evolute into ABC- and CAC-stacking under small perturbation.

In contrast to the stacking-order dependent interlayer magnetic interaction observed in bilayer CrI$_3$ \cite{sivadas2018stacking, chen2019direct}, interlayer exchange interaction in 3-SL MBT remains antiferromagnetic for all stacking configurations \cite{SM}. However, the electronic band gaps are dramatically altered by lateral shift, as depicted in \ref{fig2}(a), which even leads to a topological phase transition between quantum anomalous Hall (QAH) insulators and trivial magnetic insulators. The QAH phase with Chern number $C = 1$ is predominantly found in the vicinity of ABC-stacking, which is consistent with previous studies \cite{li2019intrinsic, otrokov2019unique}. Other stacking orders, such as CAC-stacking, result in topologically trivial magnetic insulators with $C = 0$.

Similar to sliding ferroelectricity experimentally observed in 2D van der Waals materials, the nudged elastic band (NEB) method is commonly employed to accurately calculate the transition barrier between two metastable phases. In this study,  the NEB method is utilized to determine the transition barrier by constructing an interpolated path that precisely follows the high-symmetry line [1-10]. As illustrated in \ref{fig2}(b), the energy barrier for the interlayer shift is measured to be 56 meV, close to the theoretically predicted lateral sliding energy barrier observed in CrI$_3$ \cite{sivadas2018stacking}. The moderate energy barrier for lateral displacement suggests that in-plane sliding can be employed to induce topological phase transitions in multi-SL MBT films, providing an additional degree of manipulation beyond strain, pressure and external electromagnetic field. The strong coupling between stacking order and electronic properties in MBT films provides additional degree of freedom in materials engineering, enables the manipulation of real-space patterns with chiral edge states and facilitates the exploration of the interplay between orbital and intrinsic magnetism in the moire patterns formed by twisted MBT multi-SL films \cite{lian2020flat}.

Furthermore, the quantized anomalous Hall conductance displayed in  \ref{fig2}(c), where $\sigma_{xy}=-e^2/h$ ($\sigma_{xy}=0$) is clearly observed within the band gap for ABC (CAC)-stacking, indicates a Chern number $C=1$ ($C=0$). \ref{fig2}(d)(e) illustrates the distribution of Berry curvature ($\mathbf{\Omega}$) in the momentum space for ABC- and CAC-stacking, respectively. In both stacking configurations, the Berry curvature is predominantly concentrated around the Brillouin-zone center. However, due to the coexistence of spatial inversion symmetry $\mathcal{P}$ and threefold rotation $C_{3z}$ in ABC-stacking, the Berry curvature exhibits a sixfold rotational symmetry around $\Gamma$. Conversely, CAC-stacking breaks $\mathcal{P}$, resulting in the Berry curvature only displaying a threefold rotational symmetry around $\Gamma$. The alternating positive and negative Berry curvatures in CAC-stacking renders it topologically trivial.

\begin{figure}[htbp]
	\includegraphics[width=0.9\linewidth]{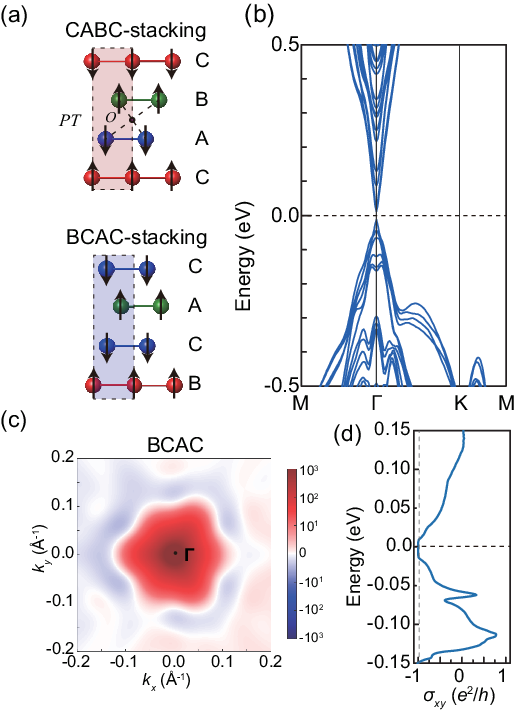}
	\caption{Electronic and topological properties of 4-SL MBT. (\textbf{a}) Schematic illustration of the presence (absence) of $\mathcal{PT}$ symmetry in  CABC-(BCAC-) stacking order. Electronic band structures (\textbf{b}), Berry curvature distribution in the momentum space (\textbf{c}) and anomalous Hall conductance $\sigma_{xy}$ as a function of Fermi energy (\textbf{d}) in the BCAC-stacking order.}
	\label{fig3}
\end{figure}

Let us turn to the even-SL case. The potential energy surface of 4-SL MBT is almost the same as in 3-SL, and the zero Hall plateau QAH state ($C = 0$) only exist close to the energetically favored CABC-stacking \cite{SM}. Unlike the odd-SL ABC stacking orders that preserve $\mathcal{P}$, antiferromagentic order in the even-SL ABC stacking orders break both $\mathcal{P}$ and time-reversal ($\mathcal{T}$) symmetries simultaneously. However, the combined symmetry operator $\mathcal{PT}$ is still respected (Fig. \ref{fig3} (a)). The existence of $\mathcal{PT}$ ensures double degeneracy at each band and vanishing Berry curvature throughout the Brillouin zone, resulting in a constraint of $C=0$. Contrarily, the local energy minimum BCAC-stacking breaks $\mathcal{PT}$, leading to the non-degenerate band structures (Fig. \ref{fig3}(b)) and non-vanishing Berry curvature in the whole momentum space (Figure \ref{fig3}(c)). The calculated Berry curvature in BCAC stacking is still concentrated at $\Gamma$ and exhibit threefold rotational symmetry. Anomalous Hall conductance shown in Fig. \ref{fig3}(d) is well quantized into $\sigma_{xy}=-e^2/h$ inside the bulk gap of 24 meV, also validating its nontrivial topological properties. We further calculate the interpolated path by the NEB method, found that topological phase transition and energy barrier in the 4-SL case is similar to the 3-SL case.

\begin{table}[htbp]
	\caption{Band gaps (in unit of meV) and Chern numbers ($C$) in multi-SL MBT thin films of distinct stacking orders.}
	\centering
	\begin{threeparttable}
	\setlength{\tabcolsep}{2mm}{
		\renewcommand\arraystretch{1.4}{
			\begin{tabular}{c|cc|cc}
			\hline\hline
			\multirow{2}{*}{SL} & \multicolumn{2}{c|}{ABC} & \multicolumn{2}{c}{CAC} \\
			\cline{2-5}
			& Band gap & $C$ & Band gap & $C$ \\
			\midrule[0.7pt]
			3  &  35  &  1  &  36  &  0  \\
			4  &  57  &  0  &  24  &  1  \\
			5  &  55  &  1  &  27  &  0  \\
			6  &  59  &  0  &  34  &  1  \\
			7  &  41  &  1  &  23  &  0  \\
			\hline\hline
	\end{tabular}}}
	\end{threeparttable}
\label{table2}
\end{table}

\begin{figure}[htbp]
	\includegraphics[width=\linewidth]{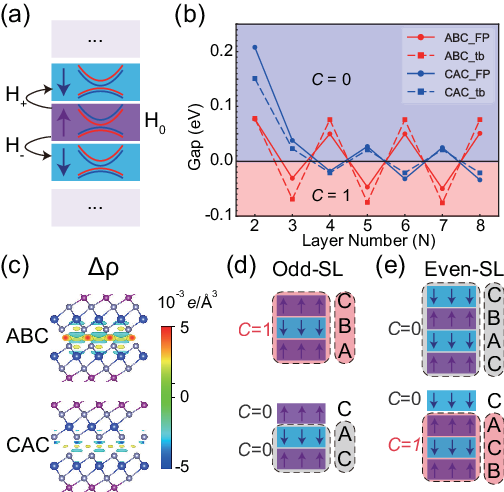}
	\caption{Underlying mechanism of stacking-dependent topology in multi-SL MBT films. (\textbf{a}) The intralayer and interlayer hopping matrix in the four-band model hamiltonians. (\textbf{b}) Comparison of topological invariants ($C$) and band gaps between first-principles calculations and model hamiltonians. (\textbf{c}) Isosurface of charge density difference ($\Delta{\rho}$) in the CAC- and ABC-stacking orders, respectively.  (\textbf{d}) Schematic illustrations of topological phases in distinct stacking orders of 3-SL and 4-SL MBT films, respectively.}
	\label{fig4}
\end{figure}

Our further calculations on other multi-SL MBT films, ranging from 5 to 7 SLs, yield similar results (Table \ref{table2}), indicating that  stack-dependent electronic and topological properties discussed above are universal. A significant band gap ($>$ 20 meV) is present in all these high-symmetry stacking orders, and topological phase transitions consistently occur in the transition path between the ABC- and CAC-stacking orders. The ABC-stacking order, which is commonly observed in multi-SL MBT films, can exhibit a QAH insulating phase with $C = 1$ in odd-SL and a zero plateau QAH phase with $C = 0$ in even-SL, consistent with previous theoretical and experimental studies \cite{li2019intrinsic, otrokov2019unique, deng2020quantum}. However, the thickness-dependent topology shows an opposite trend in the CAC-stacking order: even-SL exhibits a $C = 1$ phase with compensated antiferromagnetic (AFM) phases, while odd-SL exhibits a $C = 0$ phase with uncompensated AFM phases.

\textit{Model Analysis.---} By taking each SL as a building blocking, and assuming only nearest SLs have interlayer coupling, the minimal tight-binding hamiltonians can be 

\begin{equation}
	\begin{aligned}
		H(\mathbf{k}_\parallel) = &\sum_{\mathbf{k}, ij}\left[(H_{0}(\mathbf{k})+ H_Z)\delta_{i, j} \right.\\
		&+ \left. H_+(\mathbf{k})\delta_{i,j-1} + H_-(\mathbf{k}) \delta_{i, j+1}\right],
	\end{aligned}
\end{equation}
where $i$ denote the $i$th-SL, $\mathbf{k}_\parallel$ is the in-plane momenta ($k_x, k_y$), $H_0$, $H_{+}$ ($H_{-}$), and $H_Z$ represent intralayer, interlayer and Zeeman hamiltonians, respectively. Only the intralayer term $H_0$ is assumed to be momentum-dependent, and the latter two terms only dependent on relative lateral shift, shown as in Fig. \ref{fig4}(a). Following the effective $\vec{\mathbf{k}}\cdot\vec{\mathbf{p}}$ hamiltonians based on four low-lying state $|P1_z^{+}, \uparrow(\downarrow)\rangle$ and $|P2_z^{-}, \uparrow(\downarrow)\rangle$ \cite{zhang2009topological, liu2010model}, the intralayer and interlayer hamiltonians can be written as

\begin{equation}
	\begin{aligned}
		H_{0} &= \left[\begin{array}{cccc}
			\bar{C} + \bar{M}  &  0  &  0  &  \bar{A}_-  \\
			0  &  \bar{C} + \bar{M}  &  \bar{A}_+  &  0  \\
			0  &  \bar{A}_-  &  \bar{C} - \bar{M}  &  0  \\
			\bar{A}_+  &  0  &  0  &  \bar{C} - \bar{M}  \\
		\end{array}\right], \\
		H_{+} &= \left[\begin{array}{cccc}
			-\bar{C}_1 - \bar{M}_1  &  0  &  i \bar{B}_0  &  0  \\
			0  &  -\bar{C}_1 - \bar{M}_1  &  0  &  -i \bar{B}_0  \\
			i \bar{B}_0  &  0  &  -\bar{C}_1 + \bar{M}_1  &  0  \\
			0  &  -i \bar{B}_0  &  0  &  -\bar{C}_1 + \bar{M}_1  \\
		\end{array}\right], \\
		H_{Z} &= \left[\begin{array}{cccc}
			Z_1+Z_2  &  0  &  0  &  0  \\
			0  &  -Z_1-Z_2  &  0  &  0  \\
			0  &  0  &  Z_1-Z_2   &  0  \\
			0  &  0  &  0  &  -Z_1 + Z_2 \\ 
		\end{array}\right], 
	\end{aligned}
\end{equation}
where $\bar{C}= C_0 + C_2 (k_x^2 + k_y^2)$, $\bar{M}= M_0 + M_2 (k_x^2 + k_y^2)$, $H_{-} = H_{+}^\dagger$, $\bar{A}_\pm = A_0 (k_x \pm i k_y)$, and higher-order terms are ignored. All the parameters in the tight-binding model hamiltonians can be determined by fitting with first-principles band structures of ABC- and ACB-stacking ferromagnetic bulk MBT, whose numerical values are shown in the supplementary material \cite{SM}. Compared to ABC stacking, the interlayer hopping strength is smaller in ACB-stacking, which originates from the enlarged interlayer Te-Te distance. Multi-layer slabs with varied thickness and distinct stacking orders are built based on the above tight-binding Hamiltonians, and the corresponding electronic band gaps and topological properties are obtained, which is well in accordance with our previous first-principles calculations (Fig. \ref{fig4}(b)), confirming the validity of the aforementioned four-band model.

Given that most physical properties in density functional theory are based on charge distribution ($\rho$) in the ground state, we further performed a systemic study on the impact of lateral shift on charge density difference ($\Delta\rho$). The charge density difference ($\Delta\rho$) is defined as the net charge distribution that arises from the interaction between two subsystems, given by the formula: $\Delta\rho=\rho_{\mathrm{total}}-\rho_{\mathrm{top}}-\rho_{\mathrm{bot}}$, where $\rho_{\mathrm{total}}$ is the charge density of the combined system and $\rho_\mathrm{top}$ and $\rho_\mathrm{bot}$ are the charge densities of the non-interacting individual subsystems, the top 1-SL and remaining part of multi-SL thin films, respectively. Remarkably, $\Delta\rho$ is negligible in the CAC-stacking, but it becomes pronounced in the ABC-stacking (Fig. \ref{fig4}(c)), which means lateral shift significantly alters Te-Te interaction at the vdW interface. Compared to the consistently strong interlayer interaction in the ABC stacking, the CAC-stacking can be regarded as the superposition of two loosely connected components --- the top SL and other SLs. where interlayer interaction between these two components is relatively weaker. Thus, the CAC-stacking films exhibit a completely opposite oscillating pattern in topological properties: QAH insulating phase with $C=1$ in even SLs and zero-plateau QAH phase with $C=0$ in odd SLs (Fig. \ref{fig4}(d)).

Lateral displacement in multi-layer MBT films induce distinct types of magnetic space groups in the high-symmetry structure of odd- and even-SL MBT films, and we proposed two feasible and nondestructive optical measurement setups to determine its stacking order before conducting transports experiments. Second harmonic generation (SHG) is forbidden by inversion $\mathcal{P}$ in the ABC-stacking odd-SL MBT, but become allowed in other stacking orders, exhibiting sixfold symmetry under incident linear polarized laser \cite{chen2022basic}. Magneto-optical Kerr effect is forbidden by $\mathcal{PT}$ in the ABC-stacking even-SL MBT, and the angle-independent MOKE can be expected in other stacking orders \cite{sivadas2016gate}. The breaking of $\mathcal{P}$ and $\mathcal{PT}$ symmetries in other low-symmetry structures result in the coexistence of these two optical signals. Therefore, these two optical setups, both sensitive to material symmetries, can serve as a powerful tool to detect and distinguish local structures formed by distinct stacking orders.

\textit{Conclusion.---} In the work, we make theoretical predictions the stack-order dependent electronic and topological properties in multi-SL MnBi$_2$Te$_4$ (MBT) thin films. Lateral displacement of the topmost layer in multi-layer MBT thin films can lead to significant modifications in electronic band structures, and even leading to topological phase transitions. When the top SL is shifted from ABC stacking to CAC stacking, there emerges the opposite thickness-dependent topology contrary to the usual ABC-stacking order: uncompensated AFM trivial insulators with $C=0$ in odd SLs and compensated AFM QAH insulating insulators with $C=1$ in even SLs. Our effective tight-binding hamiltonians provide a thorough and detailed description of both ABC and CAC stacking configurations, highlighting the crucial role played by the varying interlayer Te-Te interaction. Finally, two complementary experimental setups are proposed to distinguishing different local structures. Our theoretical predictions on strong coupling between stacking order and topology emphasizes the profound effects of interlayer sliding in magnetic topological materials, opening up new possibilities for discovering a rich landscape of opportunities for tailoring and optimizing topological quantum materials.

\begin{acknowledgments}
\textit{Acknowledgments.---}This work was supported by the National Key R\&D Program of China (Grant No. 2022YFA1403800, 2023YFA1607401), the National Natural Science Foundation of China (Grant No.12274436, 11925408, 11921004), the Science Center of the National Natural Science Foundation of China (Grant No. 12188101) and  H.W. acknowledge support from the Informatization Plan of the Chinese Academy of Sciences (CASWX2021SF-0102) and the New Cornerstone Science Foundation through the XPLORER PRIZE. J. L. acknowledge support from China National Postdoctoral Program for Innovation Talents (Grant No. BX20220334).
\end{acknowledgments}

\textit{Data availability.---} The data that support the findings of the study can be found in the manuscript’s main text and Supplementary Information, and additional data is available from the corresponding author upon request.

\bibliography{reference.bib}

\begin{thebibliography}{45}%
\makeatletter
\providecommand \@ifxundefined [1]{%
 \@ifx{#1\undefined}
}%
\providecommand \@ifnum [1]{%
 \ifnum #1\expandafter \@firstoftwo
 \else \expandafter \@secondoftwo
 \fi
}%
\providecommand \@ifx [1]{%
 \ifx #1\expandafter \@firstoftwo
 \else \expandafter \@secondoftwo
 \fi
}%
\providecommand \natexlab [1]{#1}%
\providecommand \enquote  [1]{``#1''}%
\providecommand \bibnamefont  [1]{#1}%
\providecommand \bibfnamefont [1]{#1}%
\providecommand \citenamefont [1]{#1}%
\providecommand \href@noop [0]{\@secondoftwo}%
\providecommand \href [0]{\begingroup \@sanitize@url \@href}%
\providecommand \@href[1]{\@@startlink{#1}\@@href}%
\providecommand \@@href[1]{\endgroup#1\@@endlink}%
\providecommand \@sanitize@url [0]{\catcode `\\12\catcode `\$12\catcode
  `\&12\catcode `\#12\catcode `\^12\catcode `\_12\catcode `\%12\relax}%
\providecommand \@@startlink[1]{}%
\providecommand \@@endlink[0]{}%
\providecommand \url  [0]{\begingroup\@sanitize@url \@url }%
\providecommand \@url [1]{\endgroup\@href {#1}{\urlprefix }}%
\providecommand \urlprefix  [0]{URL }%
\providecommand \Eprint [0]{\href }%
\providecommand \doibase [0]{https://doi.org/}%
\providecommand \selectlanguage [0]{\@gobble}%
\providecommand \bibinfo  [0]{\@secondoftwo}%
\providecommand \bibfield  [0]{\@secondoftwo}%
\providecommand \translation [1]{[#1]}%
\providecommand \BibitemOpen [0]{}%
\providecommand \bibitemStop [0]{}%
\providecommand \bibitemNoStop [0]{.\EOS\space}%
\providecommand \EOS [0]{\spacefactor3000\relax}%
\providecommand \BibitemShut  [1]{\csname bibitem#1\endcsname}%
\let\auto@bib@innerbib\@empty
\bibitem [{\citenamefont {Wu}\ and\ \citenamefont {Li}(2021)}]{wu2021sliding}%
  \BibitemOpen
  \bibfield  {author} {\bibinfo {author} {\bibfnamefont {M.}~\bibnamefont
  {Wu}}\ and\ \bibinfo {author} {\bibfnamefont {J.}~\bibnamefont {Li}},\
  }\bibfield  {title} {\bibinfo {title} {Sliding ferroelectricity in {2D van
  der Waals} materials: Related physics and future opportunities},\ }\href
  {https://doi.org/10.1073/pnas.2115703118} {\bibfield  {journal} {\bibinfo
  {journal} {Proc. Nat. Acad. Sci.}\ }\textbf {\bibinfo {volume} {118}},\
  \bibinfo {pages} {e2115703118} (\bibinfo {year} {2021})}\BibitemShut
  {NoStop}%
\bibitem [{\citenamefont {Stern}\ \emph {et~al.}(2021)\citenamefont {Stern},
  \citenamefont {Waschitz}, \citenamefont {Cao}, \citenamefont {Nevo},
  \citenamefont {Watanabe}, \citenamefont {Taniguchi}, \citenamefont {Sela},
  \citenamefont {Urbakh}, \citenamefont {Hod},\ and\ \citenamefont
  {Shalom}}]{stern2021interfacial}%
  \BibitemOpen
  \bibfield  {author} {\bibinfo {author} {\bibfnamefont {M.~V.}\ \bibnamefont
  {Stern}}, \bibinfo {author} {\bibfnamefont {Y.}~\bibnamefont {Waschitz}},
  \bibinfo {author} {\bibfnamefont {W.}~\bibnamefont {Cao}}, \bibinfo {author}
  {\bibfnamefont {I.}~\bibnamefont {Nevo}}, \bibinfo {author} {\bibfnamefont
  {K.}~\bibnamefont {Watanabe}}, \bibinfo {author} {\bibfnamefont
  {T.}~\bibnamefont {Taniguchi}}, \bibinfo {author} {\bibfnamefont
  {E.}~\bibnamefont {Sela}}, \bibinfo {author} {\bibfnamefont {M.}~\bibnamefont
  {Urbakh}}, \bibinfo {author} {\bibfnamefont {O.}~\bibnamefont {Hod}},\ and\
  \bibinfo {author} {\bibfnamefont {M.~B.}\ \bibnamefont {Shalom}},\ }\bibfield
   {title} {\bibinfo {title} {Interfacial ferroelectricity by {van der Waals}
  sliding},\ }\href {https://doi.org/10.1126/science.abe8177} {\bibfield
  {journal} {\bibinfo  {journal} {Science}\ }\textbf {\bibinfo {volume}
  {372}},\ \bibinfo {pages} {1462} (\bibinfo {year} {2021})}\BibitemShut
  {NoStop}%
\bibitem [{\citenamefont {Sivadas}\ \emph {et~al.}(2018)\citenamefont
  {Sivadas}, \citenamefont {Okamoto}, \citenamefont {Xu}, \citenamefont
  {Fennie},\ and\ \citenamefont {Xiao}}]{sivadas2018stacking}%
  \BibitemOpen
  \bibfield  {author} {\bibinfo {author} {\bibfnamefont {N.}~\bibnamefont
  {Sivadas}}, \bibinfo {author} {\bibfnamefont {S.}~\bibnamefont {Okamoto}},
  \bibinfo {author} {\bibfnamefont {X.}~\bibnamefont {Xu}}, \bibinfo {author}
  {\bibfnamefont {C.~J.}\ \bibnamefont {Fennie}},\ and\ \bibinfo {author}
  {\bibfnamefont {D.}~\bibnamefont {Xiao}},\ }\bibfield  {title} {\bibinfo
  {title} {Stacking-dependent magnetism in bilayer cri3},\ }\href
  {https://doi.org/10.1021/acs.nanolett.8b03321} {\bibfield  {journal}
  {\bibinfo  {journal} {Nano Lett.}\ }\textbf {\bibinfo {volume} {18}},\
  \bibinfo {pages} {7658} (\bibinfo {year} {2018})}\BibitemShut {NoStop}%
\bibitem [{\citenamefont {Chen}\ \emph {et~al.}(2019)\citenamefont {Chen},
  \citenamefont {Sun}, \citenamefont {Wang}, \citenamefont {Gu}, \citenamefont
  {Xu}, \citenamefont {Wu},\ and\ \citenamefont {Gao}}]{chen2019direct}%
  \BibitemOpen
  \bibfield  {author} {\bibinfo {author} {\bibfnamefont {W.}~\bibnamefont
  {Chen}}, \bibinfo {author} {\bibfnamefont {Z.}~\bibnamefont {Sun}}, \bibinfo
  {author} {\bibfnamefont {Z.}~\bibnamefont {Wang}}, \bibinfo {author}
  {\bibfnamefont {L.}~\bibnamefont {Gu}}, \bibinfo {author} {\bibfnamefont
  {X.}~\bibnamefont {Xu}}, \bibinfo {author} {\bibfnamefont {S.}~\bibnamefont
  {Wu}},\ and\ \bibinfo {author} {\bibfnamefont {C.}~\bibnamefont {Gao}},\
  }\bibfield  {title} {\bibinfo {title} {Direct observation of {van der Waals}
  stacking–dependent interlayer magnetism},\ }\href
  {https://doi.org/10.1126/science.aav1937} {\bibfield  {journal} {\bibinfo
  {journal} {Science}\ }\textbf {\bibinfo {volume} {366}},\ \bibinfo {pages}
  {983} (\bibinfo {year} {2019})}\BibitemShut {NoStop}%
\bibitem [{\citenamefont {Andrei}\ and\ \citenamefont
  {MacDonald}(2020)}]{andrei2020graphene}%
  \BibitemOpen
  \bibfield  {author} {\bibinfo {author} {\bibfnamefont {E.~Y.}\ \bibnamefont
  {Andrei}}\ and\ \bibinfo {author} {\bibfnamefont {A.~H.}\ \bibnamefont
  {MacDonald}},\ }\bibfield  {title} {\bibinfo {title} {Graphene bilayers with
  a twist},\ }\href {https://doi.org/10.1038/s41563-020-00840-0} {\bibfield
  {journal} {\bibinfo  {journal} {Nat. Mater.}\ }\textbf {\bibinfo {volume}
  {19}},\ \bibinfo {pages} {1265} (\bibinfo {year} {2020})}\BibitemShut
  {NoStop}%
\bibitem [{\citenamefont {Bistritzer}\ and\ \citenamefont
  {MacDonald}(2011)}]{bistritzer2011moire}%
  \BibitemOpen
  \bibfield  {author} {\bibinfo {author} {\bibfnamefont {R.}~\bibnamefont
  {Bistritzer}}\ and\ \bibinfo {author} {\bibfnamefont {A.~H.}\ \bibnamefont
  {MacDonald}},\ }\bibfield  {title} {\bibinfo {title} {Moiré bands in twisted
  double-layer graphene},\ }\href {https://doi.org/10.1073/pnas.1108174108}
  {\bibfield  {journal} {\bibinfo  {journal} {Proc. Nat. Acad. Sci.}\ }\textbf
  {\bibinfo {volume} {108}},\ \bibinfo {pages} {12233} (\bibinfo {year}
  {2011})}\BibitemShut {NoStop}%
\bibitem [{\citenamefont {Cao}\ \emph {et~al.}(2018{\natexlab{a}})\citenamefont
  {Cao}, \citenamefont {Fatemi}, \citenamefont {Demir}, \citenamefont {Fang},
  \citenamefont {Tomarken}, \citenamefont {Luo}, \citenamefont
  {Sanchez-Yamagishi}, \citenamefont {Watanabe}, \citenamefont {Taniguchi},
  \citenamefont {Kaxiras}, \citenamefont {Ashoori},\ and\ \citenamefont
  {Jarillo-Herrero}}]{cao2018correlated}%
  \BibitemOpen
  \bibfield  {author} {\bibinfo {author} {\bibfnamefont {Y.}~\bibnamefont
  {Cao}}, \bibinfo {author} {\bibfnamefont {V.}~\bibnamefont {Fatemi}},
  \bibinfo {author} {\bibfnamefont {A.}~\bibnamefont {Demir}}, \bibinfo
  {author} {\bibfnamefont {S.}~\bibnamefont {Fang}}, \bibinfo {author}
  {\bibfnamefont {S.~L.}\ \bibnamefont {Tomarken}}, \bibinfo {author}
  {\bibfnamefont {J.~Y.}\ \bibnamefont {Luo}}, \bibinfo {author} {\bibfnamefont
  {J.~D.}\ \bibnamefont {Sanchez-Yamagishi}}, \bibinfo {author} {\bibfnamefont
  {K.}~\bibnamefont {Watanabe}}, \bibinfo {author} {\bibfnamefont
  {T.}~\bibnamefont {Taniguchi}}, \bibinfo {author} {\bibfnamefont
  {E.}~\bibnamefont {Kaxiras}}, \bibinfo {author} {\bibfnamefont {R.~C.}\
  \bibnamefont {Ashoori}},\ and\ \bibinfo {author} {\bibfnamefont
  {P.}~\bibnamefont {Jarillo-Herrero}},\ }\bibfield  {title} {\bibinfo {title}
  {Correlated insulator behaviour at half-filling in magic-angle graphene
  superlattices},\ }\href {https://doi.org/10.1038/nature26154} {\bibfield
  {journal} {\bibinfo  {journal} {Nature}\ }\textbf {\bibinfo {volume} {556}},\
  \bibinfo {pages} {80} (\bibinfo {year} {2018}{\natexlab{a}})}\BibitemShut
  {NoStop}%
\bibitem [{\citenamefont {Lu}\ \emph {et~al.}(2019)\citenamefont {Lu},
  \citenamefont {Stepanov}, \citenamefont {Yang}, \citenamefont {Xie},
  \citenamefont {Aamir}, \citenamefont {Das}, \citenamefont {Urgell},
  \citenamefont {Watanabe}, \citenamefont {Taniguchi}, \citenamefont {Zhang},
  \citenamefont {Bachtold}, \citenamefont {MacDonald},\ and\ \citenamefont
  {Efetov}}]{lu2019superconductors}%
  \BibitemOpen
  \bibfield  {author} {\bibinfo {author} {\bibfnamefont {X.}~\bibnamefont
  {Lu}}, \bibinfo {author} {\bibfnamefont {P.}~\bibnamefont {Stepanov}},
  \bibinfo {author} {\bibfnamefont {W.}~\bibnamefont {Yang}}, \bibinfo {author}
  {\bibfnamefont {M.}~\bibnamefont {Xie}}, \bibinfo {author} {\bibfnamefont
  {M.~A.}\ \bibnamefont {Aamir}}, \bibinfo {author} {\bibfnamefont
  {I.}~\bibnamefont {Das}}, \bibinfo {author} {\bibfnamefont {C.}~\bibnamefont
  {Urgell}}, \bibinfo {author} {\bibfnamefont {K.}~\bibnamefont {Watanabe}},
  \bibinfo {author} {\bibfnamefont {T.}~\bibnamefont {Taniguchi}}, \bibinfo
  {author} {\bibfnamefont {G.}~\bibnamefont {Zhang}}, \bibinfo {author}
  {\bibfnamefont {A.}~\bibnamefont {Bachtold}}, \bibinfo {author}
  {\bibfnamefont {A.~H.}\ \bibnamefont {MacDonald}},\ and\ \bibinfo {author}
  {\bibfnamefont {D.~K.}\ \bibnamefont {Efetov}},\ }\bibfield  {title}
  {\bibinfo {title} {Superconductors, orbital magnets and correlated states in
  magic-angle bilayer graphene},\ }\href
  {https://doi.org/10.1038/s41586-019-1695-0} {\bibfield  {journal} {\bibinfo
  {journal} {Nature}\ }\textbf {\bibinfo {volume} {574}},\ \bibinfo {pages}
  {653} (\bibinfo {year} {2019})}\BibitemShut {NoStop}%
\bibitem [{\citenamefont {Cao}\ \emph {et~al.}(2018{\natexlab{b}})\citenamefont
  {Cao}, \citenamefont {Fatemi}, \citenamefont {Fang}, \citenamefont
  {Watanabe}, \citenamefont {Taniguchi}, \citenamefont {Kaxiras},\ and\
  \citenamefont {Jarillo-Herrero}}]{cao2018unconventional}%
  \BibitemOpen
  \bibfield  {author} {\bibinfo {author} {\bibfnamefont {Y.}~\bibnamefont
  {Cao}}, \bibinfo {author} {\bibfnamefont {V.}~\bibnamefont {Fatemi}},
  \bibinfo {author} {\bibfnamefont {S.}~\bibnamefont {Fang}}, \bibinfo {author}
  {\bibfnamefont {K.}~\bibnamefont {Watanabe}}, \bibinfo {author}
  {\bibfnamefont {T.}~\bibnamefont {Taniguchi}}, \bibinfo {author}
  {\bibfnamefont {E.}~\bibnamefont {Kaxiras}},\ and\ \bibinfo {author}
  {\bibfnamefont {P.}~\bibnamefont {Jarillo-Herrero}},\ }\bibfield  {title}
  {\bibinfo {title} {Unconventional superconductivity in magic-angle graphene
  superlattices},\ }\href {https://doi.org/10.1038/nature26160} {\bibfield
  {journal} {\bibinfo  {journal} {Nature}\ }\textbf {\bibinfo {volume} {556}},\
  \bibinfo {pages} {43} (\bibinfo {year} {2018}{\natexlab{b}})}\BibitemShut
  {NoStop}%
\bibitem [{\citenamefont {Oh}\ \emph {et~al.}(2021)\citenamefont {Oh},
  \citenamefont {Nuckolls}, \citenamefont {Wong}, \citenamefont {Lee},
  \citenamefont {Liu}, \citenamefont {Watanabe}, \citenamefont {Taniguchi},\
  and\ \citenamefont {Yazdani}}]{oh2021evidence}%
  \BibitemOpen
  \bibfield  {author} {\bibinfo {author} {\bibfnamefont {M.}~\bibnamefont
  {Oh}}, \bibinfo {author} {\bibfnamefont {K.~P.}\ \bibnamefont {Nuckolls}},
  \bibinfo {author} {\bibfnamefont {D.}~\bibnamefont {Wong}}, \bibinfo {author}
  {\bibfnamefont {R.~L.}\ \bibnamefont {Lee}}, \bibinfo {author} {\bibfnamefont
  {X.}~\bibnamefont {Liu}}, \bibinfo {author} {\bibfnamefont {K.}~\bibnamefont
  {Watanabe}}, \bibinfo {author} {\bibfnamefont {T.}~\bibnamefont
  {Taniguchi}},\ and\ \bibinfo {author} {\bibfnamefont {A.}~\bibnamefont
  {Yazdani}},\ }\bibfield  {title} {\bibinfo {title} {Evidence for
  unconventional superconductivity in twisted bilayer graphene},\ }\href
  {https://doi.org/10.1038/s41586-021-04121-x} {\bibfield  {journal} {\bibinfo
  {journal} {Nature}\ }\textbf {\bibinfo {volume} {600}},\ \bibinfo {pages}
  {240} (\bibinfo {year} {2021})}\BibitemShut {NoStop}%
\bibitem [{\citenamefont {Serlin}\ \emph {et~al.}(2020)\citenamefont {Serlin},
  \citenamefont {Tschirhart}, \citenamefont {Polshyn}, \citenamefont {Zhang},
  \citenamefont {Zhu}, \citenamefont {Watanabe}, \citenamefont {Taniguchi},
  \citenamefont {Balents},\ and\ \citenamefont {Young}}]{serlin2020intrinsic}%
  \BibitemOpen
  \bibfield  {author} {\bibinfo {author} {\bibfnamefont {M.}~\bibnamefont
  {Serlin}}, \bibinfo {author} {\bibfnamefont {C.~L.}\ \bibnamefont
  {Tschirhart}}, \bibinfo {author} {\bibfnamefont {H.}~\bibnamefont {Polshyn}},
  \bibinfo {author} {\bibfnamefont {Y.}~\bibnamefont {Zhang}}, \bibinfo
  {author} {\bibfnamefont {J.}~\bibnamefont {Zhu}}, \bibinfo {author}
  {\bibfnamefont {K.}~\bibnamefont {Watanabe}}, \bibinfo {author}
  {\bibfnamefont {T.}~\bibnamefont {Taniguchi}}, \bibinfo {author}
  {\bibfnamefont {L.}~\bibnamefont {Balents}},\ and\ \bibinfo {author}
  {\bibfnamefont {A.~F.}\ \bibnamefont {Young}},\ }\bibfield  {title} {\bibinfo
  {title} {Intrinsic quantized anomalous hall effect in a moiré
  heterostructure},\ }\href {https://doi.org/10.1126/science.aay5533}
  {\bibfield  {journal} {\bibinfo  {journal} {Science}\ }\textbf {\bibinfo
  {volume} {367}},\ \bibinfo {pages} {900} (\bibinfo {year}
  {2020})}\BibitemShut {NoStop}%
\bibitem [{\citenamefont {Li}\ \emph {et~al.}(2019{\natexlab{a}})\citenamefont
  {Li}, \citenamefont {Jiang}, \citenamefont {Sivadas}, \citenamefont {Wang},
  \citenamefont {Xu}, \citenamefont {Weber}, \citenamefont {Goldberger},
  \citenamefont {Watanabe}, \citenamefont {Taniguchi}, \citenamefont {Fennie},
  \citenamefont {Fai~Mak},\ and\ \citenamefont {Shan}}]{li2019pressure}%
  \BibitemOpen
  \bibfield  {author} {\bibinfo {author} {\bibfnamefont {T.}~\bibnamefont
  {Li}}, \bibinfo {author} {\bibfnamefont {S.}~\bibnamefont {Jiang}}, \bibinfo
  {author} {\bibfnamefont {N.}~\bibnamefont {Sivadas}}, \bibinfo {author}
  {\bibfnamefont {Z.}~\bibnamefont {Wang}}, \bibinfo {author} {\bibfnamefont
  {Y.}~\bibnamefont {Xu}}, \bibinfo {author} {\bibfnamefont {D.}~\bibnamefont
  {Weber}}, \bibinfo {author} {\bibfnamefont {J.~E.}\ \bibnamefont
  {Goldberger}}, \bibinfo {author} {\bibfnamefont {K.}~\bibnamefont
  {Watanabe}}, \bibinfo {author} {\bibfnamefont {T.}~\bibnamefont {Taniguchi}},
  \bibinfo {author} {\bibfnamefont {C.~J.}\ \bibnamefont {Fennie}}, \bibinfo
  {author} {\bibfnamefont {K.}~\bibnamefont {Fai~Mak}},\ and\ \bibinfo {author}
  {\bibfnamefont {J.}~\bibnamefont {Shan}},\ }\bibfield  {title} {\bibinfo
  {title} {Pressure-controlled interlayer magnetism in atomically thin
  {CrI$_3$}},\ }\href {https://doi.org/10.1038/s41563-019-0506-1} {\bibfield
  {journal} {\bibinfo  {journal} {Nat. Mater.}\ }\textbf {\bibinfo {volume}
  {18}},\ \bibinfo {pages} {1303} (\bibinfo {year}
  {2019}{\natexlab{a}})}\BibitemShut {NoStop}%
\bibitem [{\citenamefont {Song}\ \emph {et~al.}(2019)\citenamefont {Song},
  \citenamefont {Fei}, \citenamefont {Yankowitz}, \citenamefont {Lin},
  \citenamefont {Jiang}, \citenamefont {Hwangbo}, \citenamefont {Zhang},
  \citenamefont {Sun}, \citenamefont {Taniguchi}, \citenamefont {Watanabe},
  \citenamefont {McGuire}, \citenamefont {Graf}, \citenamefont {Cao},
  \citenamefont {Chu}, \citenamefont {Cobden}, \citenamefont {Dean},
  \citenamefont {Xiao},\ and\ \citenamefont {Xu}}]{song2019switching}%
  \BibitemOpen
  \bibfield  {author} {\bibinfo {author} {\bibfnamefont {T.}~\bibnamefont
  {Song}}, \bibinfo {author} {\bibfnamefont {Z.}~\bibnamefont {Fei}}, \bibinfo
  {author} {\bibfnamefont {M.}~\bibnamefont {Yankowitz}}, \bibinfo {author}
  {\bibfnamefont {Z.}~\bibnamefont {Lin}}, \bibinfo {author} {\bibfnamefont
  {Q.}~\bibnamefont {Jiang}}, \bibinfo {author} {\bibfnamefont
  {K.}~\bibnamefont {Hwangbo}}, \bibinfo {author} {\bibfnamefont
  {Q.}~\bibnamefont {Zhang}}, \bibinfo {author} {\bibfnamefont
  {B.}~\bibnamefont {Sun}}, \bibinfo {author} {\bibfnamefont {T.}~\bibnamefont
  {Taniguchi}}, \bibinfo {author} {\bibfnamefont {K.}~\bibnamefont {Watanabe}},
  \bibinfo {author} {\bibfnamefont {M.~A.}\ \bibnamefont {McGuire}}, \bibinfo
  {author} {\bibfnamefont {D.}~\bibnamefont {Graf}}, \bibinfo {author}
  {\bibfnamefont {T.}~\bibnamefont {Cao}}, \bibinfo {author} {\bibfnamefont
  {J.-H.}\ \bibnamefont {Chu}}, \bibinfo {author} {\bibfnamefont {D.~H.}\
  \bibnamefont {Cobden}}, \bibinfo {author} {\bibfnamefont {C.~R.}\
  \bibnamefont {Dean}}, \bibinfo {author} {\bibfnamefont {D.}~\bibnamefont
  {Xiao}},\ and\ \bibinfo {author} {\bibfnamefont {X.}~\bibnamefont {Xu}},\
  }\bibfield  {title} {\bibinfo {title} {Switching {2D} magnetic states via
  pressure tuning of layer stacking},\ }\href
  {https://doi.org/10.1038/s41563-019-0505-2} {\bibfield  {journal} {\bibinfo
  {journal} {Nat. Mater.}\ }\textbf {\bibinfo {volume} {18}},\ \bibinfo {pages}
  {1298} (\bibinfo {year} {2019})}\BibitemShut {NoStop}%
\bibitem [{\citenamefont {Jiang}\ \emph {et~al.}(2019)\citenamefont {Jiang},
  \citenamefont {Wang}, \citenamefont {Chen}, \citenamefont {Zhong},
  \citenamefont {Yuan}, \citenamefont {Lu},\ and\ \citenamefont
  {Ji}}]{jiang2019stacking}%
  \BibitemOpen
  \bibfield  {author} {\bibinfo {author} {\bibfnamefont {P.}~\bibnamefont
  {Jiang}}, \bibinfo {author} {\bibfnamefont {C.}~\bibnamefont {Wang}},
  \bibinfo {author} {\bibfnamefont {D.}~\bibnamefont {Chen}}, \bibinfo {author}
  {\bibfnamefont {Z.}~\bibnamefont {Zhong}}, \bibinfo {author} {\bibfnamefont
  {Z.}~\bibnamefont {Yuan}}, \bibinfo {author} {\bibfnamefont {Z.-Y.}\
  \bibnamefont {Lu}},\ and\ \bibinfo {author} {\bibfnamefont {W.}~\bibnamefont
  {Ji}},\ }\bibfield  {title} {\bibinfo {title} {Stacking tunable interlayer
  magnetism in bilayer ${\mathrm{cri}}_{3}$},\ }\href
  {https://doi.org/10.1103/PhysRevB.99.144401} {\bibfield  {journal} {\bibinfo
  {journal} {Phys. Rev. B}\ }\textbf {\bibinfo {volume} {99}},\ \bibinfo
  {pages} {144401} (\bibinfo {year} {2019})}\BibitemShut {NoStop}%
\bibitem [{\citenamefont {Li}\ \emph {et~al.}(2019{\natexlab{b}})\citenamefont
  {Li}, \citenamefont {Li}, \citenamefont {Du}, \citenamefont {Wang},
  \citenamefont {Gu}, \citenamefont {Zhang}, \citenamefont {He}, \citenamefont
  {Duan},\ and\ \citenamefont {Xu}}]{li2019intrinsic}%
  \BibitemOpen
  \bibfield  {author} {\bibinfo {author} {\bibfnamefont {J.}~\bibnamefont
  {Li}}, \bibinfo {author} {\bibfnamefont {Y.}~\bibnamefont {Li}}, \bibinfo
  {author} {\bibfnamefont {S.}~\bibnamefont {Du}}, \bibinfo {author}
  {\bibfnamefont {Z.}~\bibnamefont {Wang}}, \bibinfo {author} {\bibfnamefont
  {B.-L.}\ \bibnamefont {Gu}}, \bibinfo {author} {\bibfnamefont {S.-C.}\
  \bibnamefont {Zhang}}, \bibinfo {author} {\bibfnamefont {K.}~\bibnamefont
  {He}}, \bibinfo {author} {\bibfnamefont {W.}~\bibnamefont {Duan}},\ and\
  \bibinfo {author} {\bibfnamefont {Y.}~\bibnamefont {Xu}},\ }\bibfield
  {title} {\bibinfo {title} {Intrinsic magnetic topological insulators in van
  der waals layered {MnBi$_2$Te$_4$}-family materials},\ }\bibfield  {journal}
  {\bibinfo  {journal} {Sci. Adv.}\ }\textbf {\bibinfo {volume} {5}},\ \href
  {https://doi.org/10.1126/sciadv.aaw5685} {10.1126/sciadv.aaw5685} (\bibinfo
  {year} {2019}{\natexlab{b}})\BibitemShut {NoStop}%
\bibitem [{\citenamefont {Otrokov}\ \emph
  {et~al.}(2019{\natexlab{a}})\citenamefont {Otrokov}, \citenamefont {Rusinov},
  \citenamefont {Blanco-Rey}, \citenamefont {Hoffmann}, \citenamefont
  {Vyazovskaya}, \citenamefont {Eremeev}, \citenamefont {Ernst}, \citenamefont
  {Echenique}, \citenamefont {Arnau},\ and\ \citenamefont
  {Chulkov}}]{otrokov2019unique}%
  \BibitemOpen
  \bibfield  {author} {\bibinfo {author} {\bibfnamefont {M.~M.}\ \bibnamefont
  {Otrokov}}, \bibinfo {author} {\bibfnamefont {I.~P.}\ \bibnamefont
  {Rusinov}}, \bibinfo {author} {\bibfnamefont {M.}~\bibnamefont {Blanco-Rey}},
  \bibinfo {author} {\bibfnamefont {M.}~\bibnamefont {Hoffmann}}, \bibinfo
  {author} {\bibfnamefont {A.~Y.}\ \bibnamefont {Vyazovskaya}}, \bibinfo
  {author} {\bibfnamefont {S.~V.}\ \bibnamefont {Eremeev}}, \bibinfo {author}
  {\bibfnamefont {A.}~\bibnamefont {Ernst}}, \bibinfo {author} {\bibfnamefont
  {P.~M.}\ \bibnamefont {Echenique}}, \bibinfo {author} {\bibfnamefont
  {A.}~\bibnamefont {Arnau}},\ and\ \bibinfo {author} {\bibfnamefont {E.~V.}\
  \bibnamefont {Chulkov}},\ }\bibfield  {title} {\bibinfo {title} {Unique
  thickness-dependent properties of the van der waals interlayer
  antiferromagnet {${\mathrm{MnBi}}_{2}{\mathrm{Te}}_{4}$} films},\ }\href
  {https://doi.org/10.1103/PhysRevLett.122.107202} {\bibfield  {journal}
  {\bibinfo  {journal} {Phys. Rev. Lett.}\ }\textbf {\bibinfo {volume} {122}},\
  \bibinfo {pages} {107202} (\bibinfo {year} {2019}{\natexlab{a}})}\BibitemShut
  {NoStop}%
\bibitem [{\citenamefont {Zhang}\ \emph {et~al.}(2019)\citenamefont {Zhang},
  \citenamefont {Shi}, \citenamefont {Zhu}, \citenamefont {Xing}, \citenamefont
  {Zhang},\ and\ \citenamefont {Wang}}]{zhang2019topological}%
  \BibitemOpen
  \bibfield  {author} {\bibinfo {author} {\bibfnamefont {D.}~\bibnamefont
  {Zhang}}, \bibinfo {author} {\bibfnamefont {M.}~\bibnamefont {Shi}}, \bibinfo
  {author} {\bibfnamefont {T.}~\bibnamefont {Zhu}}, \bibinfo {author}
  {\bibfnamefont {D.}~\bibnamefont {Xing}}, \bibinfo {author} {\bibfnamefont
  {H.}~\bibnamefont {Zhang}},\ and\ \bibinfo {author} {\bibfnamefont
  {J.}~\bibnamefont {Wang}},\ }\bibfield  {title} {\bibinfo {title}
  {Topological {Axion} states in the magnetic insulator {MnBi$_2$Te$_4$} with
  the quantized magnetoelectric effect},\ }\href
  {https://doi.org/10.1103/PhysRevLett.122.206401} {\bibfield  {journal}
  {\bibinfo  {journal} {Phys. Rev. Lett.}\ }\textbf {\bibinfo {volume} {122}},\
  \bibinfo {pages} {206401} (\bibinfo {year} {2019})}\BibitemShut {NoStop}%
\bibitem [{\citenamefont {Lee}\ \emph {et~al.}(2013)\citenamefont {Lee},
  \citenamefont {Kim}, \citenamefont {Park}, \citenamefont {Chung},
  \citenamefont {Lim}, \citenamefont {Seo},\ and\ \citenamefont
  {Park}}]{lee2013crystal}%
  \BibitemOpen
  \bibfield  {author} {\bibinfo {author} {\bibfnamefont {D.~S.}\ \bibnamefont
  {Lee}}, \bibinfo {author} {\bibfnamefont {T.-H.}\ \bibnamefont {Kim}},
  \bibinfo {author} {\bibfnamefont {C.-H.}\ \bibnamefont {Park}}, \bibinfo
  {author} {\bibfnamefont {C.-Y.}\ \bibnamefont {Chung}}, \bibinfo {author}
  {\bibfnamefont {Y.~S.}\ \bibnamefont {Lim}}, \bibinfo {author} {\bibfnamefont
  {W.-S.}\ \bibnamefont {Seo}},\ and\ \bibinfo {author} {\bibfnamefont {H.-H.}\
  \bibnamefont {Park}},\ }\bibfield  {title} {\bibinfo {title} {Crystal
  structure, properties and nanostructuring of a new layered chalcogenide
  semiconductor, {Bi$_2$MnTe$_4$}},\ }\href@noop {} {\bibfield  {journal}
  {\bibinfo  {journal} {Cryst. Eng. Comm.}\ }\textbf {\bibinfo {volume} {15}},\
  \bibinfo {pages} {5532} (\bibinfo {year} {2013})}\BibitemShut {NoStop}%
\bibitem [{\citenamefont {Gong}\ \emph {et~al.}(2019)\citenamefont {Gong},
  \citenamefont {Guo}, \citenamefont {Li}, \citenamefont {Zhu}, \citenamefont
  {Liao}, \citenamefont {Liu}, \citenamefont {Zhang}, \citenamefont {Gu},
  \citenamefont {Tang}, \citenamefont {Feng}, \citenamefont {Zhang},
  \citenamefont {Li}, \citenamefont {Song}, \citenamefont {Wang}, \citenamefont
  {Yu}, \citenamefont {Chen}, \citenamefont {Wang}, \citenamefont {Yao},
  \citenamefont {Duan}, \citenamefont {Xu}, \citenamefont {Zhang},
  \citenamefont {Ma}, \citenamefont {Xue},\ and\ \citenamefont
  {He}}]{gong2019experimental}%
  \BibitemOpen
  \bibfield  {author} {\bibinfo {author} {\bibfnamefont {Y.}~\bibnamefont
  {Gong}}, \bibinfo {author} {\bibfnamefont {J.}~\bibnamefont {Guo}}, \bibinfo
  {author} {\bibfnamefont {J.}~\bibnamefont {Li}}, \bibinfo {author}
  {\bibfnamefont {K.}~\bibnamefont {Zhu}}, \bibinfo {author} {\bibfnamefont
  {M.}~\bibnamefont {Liao}}, \bibinfo {author} {\bibfnamefont {X.}~\bibnamefont
  {Liu}}, \bibinfo {author} {\bibfnamefont {Q.}~\bibnamefont {Zhang}}, \bibinfo
  {author} {\bibfnamefont {L.}~\bibnamefont {Gu}}, \bibinfo {author}
  {\bibfnamefont {L.}~\bibnamefont {Tang}}, \bibinfo {author} {\bibfnamefont
  {X.}~\bibnamefont {Feng}}, \bibinfo {author} {\bibfnamefont {D.}~\bibnamefont
  {Zhang}}, \bibinfo {author} {\bibfnamefont {W.}~\bibnamefont {Li}}, \bibinfo
  {author} {\bibfnamefont {C.}~\bibnamefont {Song}}, \bibinfo {author}
  {\bibfnamefont {L.}~\bibnamefont {Wang}}, \bibinfo {author} {\bibfnamefont
  {P.}~\bibnamefont {Yu}}, \bibinfo {author} {\bibfnamefont {X.}~\bibnamefont
  {Chen}}, \bibinfo {author} {\bibfnamefont {Y.}~\bibnamefont {Wang}}, \bibinfo
  {author} {\bibfnamefont {H.}~\bibnamefont {Yao}}, \bibinfo {author}
  {\bibfnamefont {W.}~\bibnamefont {Duan}}, \bibinfo {author} {\bibfnamefont
  {Y.}~\bibnamefont {Xu}}, \bibinfo {author} {\bibfnamefont {S.-C.}\
  \bibnamefont {Zhang}}, \bibinfo {author} {\bibfnamefont {X.}~\bibnamefont
  {Ma}}, \bibinfo {author} {\bibfnamefont {Q.-K.}\ \bibnamefont {Xue}},\ and\
  \bibinfo {author} {\bibfnamefont {K.}~\bibnamefont {He}},\ }\bibfield
  {title} {\bibinfo {title} {Experimental realization of an intrinsic magnetic
  topological insulator*},\ }\href
  {https://doi.org/10.1088/0256-307X/36/7/076801} {\bibfield  {journal}
  {\bibinfo  {journal} {Chin. Phys. Lett.}\ }\textbf {\bibinfo {volume} {36}},\
  \bibinfo {pages} {076801} (\bibinfo {year} {2019})}\BibitemShut {NoStop}%
\bibitem [{\citenamefont {Otrokov}\ \emph
  {et~al.}(2019{\natexlab{b}})\citenamefont {Otrokov}, \citenamefont
  {Klimovskikh}, \citenamefont {Bentmann}, \citenamefont {Estyunin},
  \citenamefont {Zeugner}, \citenamefont {Aliev}, \citenamefont {Gaß},
  \citenamefont {Wolter}, \citenamefont {Koroleva}, \citenamefont {Shikin},
  \citenamefont {Blanco-Rey}, \citenamefont {Hoffmann}, \citenamefont
  {Rusinov}, \citenamefont {Vyazovskaya}, \citenamefont {Eremeev},
  \citenamefont {Koroteev}, \citenamefont {Kuznetsov}, \citenamefont {Freyse},
  \citenamefont {Sánchez-Barriga}, \citenamefont {Amiraslanov}, \citenamefont
  {Babanly}, \citenamefont {Mamedov}, \citenamefont {Abdullayev}, \citenamefont
  {Zverev}, \citenamefont {Alfonsov}, \citenamefont {Kataev}, \citenamefont
  {Büchner}, \citenamefont {Schwier}, \citenamefont {Kumar}, \citenamefont
  {Kimura}, \citenamefont {Petaccia}, \citenamefont {Di~Santo}, \citenamefont
  {Vidal}, \citenamefont {Schatz}, \citenamefont {Kißner}, \citenamefont
  {Ünzelmann}, \citenamefont {Min}, \citenamefont {Moser}, \citenamefont
  {Peixoto}, \citenamefont {Reinert}, \citenamefont {Ernst}, \citenamefont
  {Echenique}, \citenamefont {Isaeva},\ and\ \citenamefont
  {Chulkov}}]{otrokov2019prediction}%
  \BibitemOpen
  \bibfield  {author} {\bibinfo {author} {\bibfnamefont {M.~M.}\ \bibnamefont
  {Otrokov}}, \bibinfo {author} {\bibfnamefont {I.~I.}\ \bibnamefont
  {Klimovskikh}}, \bibinfo {author} {\bibfnamefont {H.}~\bibnamefont
  {Bentmann}}, \bibinfo {author} {\bibfnamefont {D.}~\bibnamefont {Estyunin}},
  \bibinfo {author} {\bibfnamefont {A.}~\bibnamefont {Zeugner}}, \bibinfo
  {author} {\bibfnamefont {Z.~S.}\ \bibnamefont {Aliev}}, \bibinfo {author}
  {\bibfnamefont {S.}~\bibnamefont {Gaß}}, \bibinfo {author} {\bibfnamefont
  {A.~U.~B.}\ \bibnamefont {Wolter}}, \bibinfo {author} {\bibfnamefont {A.~V.}\
  \bibnamefont {Koroleva}}, \bibinfo {author} {\bibfnamefont {A.~M.}\
  \bibnamefont {Shikin}}, \bibinfo {author} {\bibfnamefont {M.}~\bibnamefont
  {Blanco-Rey}}, \bibinfo {author} {\bibfnamefont {M.}~\bibnamefont
  {Hoffmann}}, \bibinfo {author} {\bibfnamefont {I.~P.}\ \bibnamefont
  {Rusinov}}, \bibinfo {author} {\bibfnamefont {A.~Y.}\ \bibnamefont
  {Vyazovskaya}}, \bibinfo {author} {\bibfnamefont {S.~V.}\ \bibnamefont
  {Eremeev}}, \bibinfo {author} {\bibfnamefont {Y.~M.}\ \bibnamefont
  {Koroteev}}, \bibinfo {author} {\bibfnamefont {V.~M.}\ \bibnamefont
  {Kuznetsov}}, \bibinfo {author} {\bibfnamefont {F.}~\bibnamefont {Freyse}},
  \bibinfo {author} {\bibfnamefont {J.}~\bibnamefont {Sánchez-Barriga}},
  \bibinfo {author} {\bibfnamefont {I.~R.}\ \bibnamefont {Amiraslanov}},
  \bibinfo {author} {\bibfnamefont {M.~B.}\ \bibnamefont {Babanly}}, \bibinfo
  {author} {\bibfnamefont {N.~T.}\ \bibnamefont {Mamedov}}, \bibinfo {author}
  {\bibfnamefont {N.~A.}\ \bibnamefont {Abdullayev}}, \bibinfo {author}
  {\bibfnamefont {V.~N.}\ \bibnamefont {Zverev}}, \bibinfo {author}
  {\bibfnamefont {A.}~\bibnamefont {Alfonsov}}, \bibinfo {author}
  {\bibfnamefont {V.}~\bibnamefont {Kataev}}, \bibinfo {author} {\bibfnamefont
  {B.}~\bibnamefont {Büchner}}, \bibinfo {author} {\bibfnamefont {E.~F.}\
  \bibnamefont {Schwier}}, \bibinfo {author} {\bibfnamefont {S.}~\bibnamefont
  {Kumar}}, \bibinfo {author} {\bibfnamefont {A.}~\bibnamefont {Kimura}},
  \bibinfo {author} {\bibfnamefont {L.}~\bibnamefont {Petaccia}}, \bibinfo
  {author} {\bibfnamefont {G.}~\bibnamefont {Di~Santo}}, \bibinfo {author}
  {\bibfnamefont {R.~C.}\ \bibnamefont {Vidal}}, \bibinfo {author}
  {\bibfnamefont {S.}~\bibnamefont {Schatz}}, \bibinfo {author} {\bibfnamefont
  {K.}~\bibnamefont {Kißner}}, \bibinfo {author} {\bibfnamefont
  {M.}~\bibnamefont {Ünzelmann}}, \bibinfo {author} {\bibfnamefont {C.~H.}\
  \bibnamefont {Min}}, \bibinfo {author} {\bibfnamefont {S.}~\bibnamefont
  {Moser}}, \bibinfo {author} {\bibfnamefont {T.~R.~F.}\ \bibnamefont
  {Peixoto}}, \bibinfo {author} {\bibfnamefont {F.}~\bibnamefont {Reinert}},
  \bibinfo {author} {\bibfnamefont {A.}~\bibnamefont {Ernst}}, \bibinfo
  {author} {\bibfnamefont {P.~M.}\ \bibnamefont {Echenique}}, \bibinfo {author}
  {\bibfnamefont {A.}~\bibnamefont {Isaeva}},\ and\ \bibinfo {author}
  {\bibfnamefont {E.~V.}\ \bibnamefont {Chulkov}},\ }\bibfield  {title}
  {\bibinfo {title} {Prediction and observation of an antiferromagnetic
  topological insulator},\ }\href {https://doi.org/10.1038/s41586-019-1840-9}
  {\bibfield  {journal} {\bibinfo  {journal} {Nature}\ }\textbf {\bibinfo
  {volume} {576}},\ \bibinfo {pages} {416} (\bibinfo {year}
  {2019}{\natexlab{b}})}\BibitemShut {NoStop}%
\bibitem [{\citenamefont {Yan}\ \emph {et~al.}(2019)\citenamefont {Yan},
  \citenamefont {Zhang}, \citenamefont {Heitmann}, \citenamefont {Huang},
  \citenamefont {Chen}, \citenamefont {Cheng}, \citenamefont {Wu},
  \citenamefont {Vaknin}, \citenamefont {Sales},\ and\ \citenamefont
  {McQueeney}}]{yan2019crystal}%
  \BibitemOpen
  \bibfield  {author} {\bibinfo {author} {\bibfnamefont {J.-Q.}\ \bibnamefont
  {Yan}}, \bibinfo {author} {\bibfnamefont {Q.}~\bibnamefont {Zhang}}, \bibinfo
  {author} {\bibfnamefont {T.}~\bibnamefont {Heitmann}}, \bibinfo {author}
  {\bibfnamefont {Z.}~\bibnamefont {Huang}}, \bibinfo {author} {\bibfnamefont
  {K.~Y.}\ \bibnamefont {Chen}}, \bibinfo {author} {\bibfnamefont {J.-G.}\
  \bibnamefont {Cheng}}, \bibinfo {author} {\bibfnamefont {W.}~\bibnamefont
  {Wu}}, \bibinfo {author} {\bibfnamefont {D.}~\bibnamefont {Vaknin}}, \bibinfo
  {author} {\bibfnamefont {B.~C.}\ \bibnamefont {Sales}},\ and\ \bibinfo
  {author} {\bibfnamefont {R.~J.}\ \bibnamefont {McQueeney}},\ }\bibfield
  {title} {\bibinfo {title} {Crystal growth and magnetic structure of
  {${\mathrm{MnBi}}_{2}{\mathrm{Te}}_{4}$}},\ }\href
  {https://doi.org/10.1103/PhysRevMaterials.3.064202} {\bibfield  {journal}
  {\bibinfo  {journal} {Phys. Rev. Mater.}\ }\textbf {\bibinfo {volume} {3}},\
  \bibinfo {pages} {064202} (\bibinfo {year} {2019})}\BibitemShut {NoStop}%
\bibitem [{\citenamefont {Cui}\ \emph {et~al.}(2019)\citenamefont {Cui},
  \citenamefont {Shi}, \citenamefont {Wang}, \citenamefont {Yu}, \citenamefont
  {Wu}, \citenamefont {Luo}, \citenamefont {Ying},\ and\ \citenamefont
  {Chen}}]{cui2019transport}%
  \BibitemOpen
  \bibfield  {author} {\bibinfo {author} {\bibfnamefont {J.}~\bibnamefont
  {Cui}}, \bibinfo {author} {\bibfnamefont {M.}~\bibnamefont {Shi}}, \bibinfo
  {author} {\bibfnamefont {H.}~\bibnamefont {Wang}}, \bibinfo {author}
  {\bibfnamefont {F.}~\bibnamefont {Yu}}, \bibinfo {author} {\bibfnamefont
  {T.}~\bibnamefont {Wu}}, \bibinfo {author} {\bibfnamefont {X.}~\bibnamefont
  {Luo}}, \bibinfo {author} {\bibfnamefont {J.}~\bibnamefont {Ying}},\ and\
  \bibinfo {author} {\bibfnamefont {X.}~\bibnamefont {Chen}},\ }\bibfield
  {title} {\bibinfo {title} {Transport properties of thin flakes of the
  antiferromagnetic topological insulator
  $\mathrm{MnB}{\mathrm{i}}_{2}\mathrm{T}{\mathrm{e}}_{4}$},\ }\href
  {https://doi.org/10.1103/PhysRevB.99.155125} {\bibfield  {journal} {\bibinfo
  {journal} {Phys. Rev. B}\ }\textbf {\bibinfo {volume} {99}},\ \bibinfo
  {pages} {155125} (\bibinfo {year} {2019})}\BibitemShut {NoStop}%
\bibitem [{\citenamefont {Li}\ \emph {et~al.}(2020)\citenamefont {Li},
  \citenamefont {Liu}, \citenamefont {Liu}, \citenamefont {Zhang},
  \citenamefont {Xu}, \citenamefont {Yu}, \citenamefont {Wu}, \citenamefont
  {Zhang},\ and\ \citenamefont {Fan}}]{li2020antiferromagnetic}%
  \BibitemOpen
  \bibfield  {author} {\bibinfo {author} {\bibfnamefont {H.}~\bibnamefont
  {Li}}, \bibinfo {author} {\bibfnamefont {S.}~\bibnamefont {Liu}}, \bibinfo
  {author} {\bibfnamefont {C.}~\bibnamefont {Liu}}, \bibinfo {author}
  {\bibfnamefont {J.}~\bibnamefont {Zhang}}, \bibinfo {author} {\bibfnamefont
  {Y.}~\bibnamefont {Xu}}, \bibinfo {author} {\bibfnamefont {R.}~\bibnamefont
  {Yu}}, \bibinfo {author} {\bibfnamefont {Y.}~\bibnamefont {Wu}}, \bibinfo
  {author} {\bibfnamefont {Y.}~\bibnamefont {Zhang}},\ and\ \bibinfo {author}
  {\bibfnamefont {S.}~\bibnamefont {Fan}},\ }\bibfield  {title} {\bibinfo
  {title} {Antiferromagnetic topological insulator {MnBi$_2$Te$_4$}: synthesis
  and magnetic properties},\ }\href {https://doi.org/10.1039/C9CP05634C}
  {\bibfield  {journal} {\bibinfo  {journal} {Phys. Chem. Chem. Phys.}\
  }\textbf {\bibinfo {volume} {22}},\ \bibinfo {pages} {556} (\bibinfo {year}
  {2020})}\BibitemShut {NoStop}%
\bibitem [{\citenamefont {Mong}\ \emph {et~al.}(2010)\citenamefont {Mong},
  \citenamefont {Essin},\ and\ \citenamefont
  {Moore}}]{mong2010antiferromagnetic}%
  \BibitemOpen
  \bibfield  {author} {\bibinfo {author} {\bibfnamefont {R.~S.}\ \bibnamefont
  {Mong}}, \bibinfo {author} {\bibfnamefont {A.~M.}\ \bibnamefont {Essin}},\
  and\ \bibinfo {author} {\bibfnamefont {J.~E.}\ \bibnamefont {Moore}},\
  }\bibfield  {title} {\bibinfo {title} {Antiferromagnetic topological
  insulators},\ }\href@noop {} {\bibfield  {journal} {\bibinfo  {journal}
  {Phys. Rev. B}\ }\textbf {\bibinfo {volume} {81}},\ \bibinfo {pages} {245209}
  (\bibinfo {year} {2010})}\BibitemShut {NoStop}%
\bibitem [{\citenamefont {Li}\ \emph {et~al.}(2023)\citenamefont {Li},
  \citenamefont {Liu}, \citenamefont {Liu}, \citenamefont {Wang}, \citenamefont
  {Lu},\ and\ \citenamefont {Xie}}]{li2023progress}%
  \BibitemOpen
  \bibfield  {author} {\bibinfo {author} {\bibfnamefont {S.}~\bibnamefont
  {Li}}, \bibinfo {author} {\bibfnamefont {T.}~\bibnamefont {Liu}}, \bibinfo
  {author} {\bibfnamefont {C.}~\bibnamefont {Liu}}, \bibinfo {author}
  {\bibfnamefont {Y.}~\bibnamefont {Wang}}, \bibinfo {author} {\bibfnamefont
  {H.-Z.}\ \bibnamefont {Lu}},\ and\ \bibinfo {author} {\bibfnamefont {X.~C.}\
  \bibnamefont {Xie}},\ }\bibfield  {title} {\bibinfo {title} {Progress on the
  antiferromagnetic topological insulator {MnBi$_2$Te$_4$}},\ }\href
  {https://doi.org/10.1093/nsr/nwac296} {\bibfield  {journal} {\bibinfo
  {journal} {Natl. Sci. Rev.}\ }\textbf {\bibinfo {volume} {11}},\ \bibinfo
  {pages} {nwac296} (\bibinfo {year} {2023})}\BibitemShut {NoStop}%
\bibitem [{\citenamefont {Wang}\ \emph {et~al.}(2023)\citenamefont {Wang},
  \citenamefont {Ma}, \citenamefont {Hao}, \citenamefont {Cai}, \citenamefont
  {Rong}, \citenamefont {Zhang}, \citenamefont {Chen}, \citenamefont {Zhang},
  \citenamefont {Lin}, \citenamefont {Zhao}, \citenamefont {Liu}, \citenamefont
  {Liu},\ and\ \citenamefont {Chen}}]{wang2023topological}%
  \BibitemOpen
  \bibfield  {author} {\bibinfo {author} {\bibfnamefont {Y.}~\bibnamefont
  {Wang}}, \bibinfo {author} {\bibfnamefont {X.-M.}\ \bibnamefont {Ma}},
  \bibinfo {author} {\bibfnamefont {Z.}~\bibnamefont {Hao}}, \bibinfo {author}
  {\bibfnamefont {Y.}~\bibnamefont {Cai}}, \bibinfo {author} {\bibfnamefont
  {H.}~\bibnamefont {Rong}}, \bibinfo {author} {\bibfnamefont {F.}~\bibnamefont
  {Zhang}}, \bibinfo {author} {\bibfnamefont {W.}~\bibnamefont {Chen}},
  \bibinfo {author} {\bibfnamefont {C.}~\bibnamefont {Zhang}}, \bibinfo
  {author} {\bibfnamefont {J.}~\bibnamefont {Lin}}, \bibinfo {author}
  {\bibfnamefont {Y.}~\bibnamefont {Zhao}}, \bibinfo {author} {\bibfnamefont
  {C.}~\bibnamefont {Liu}}, \bibinfo {author} {\bibfnamefont {Q.}~\bibnamefont
  {Liu}},\ and\ \bibinfo {author} {\bibfnamefont {C.}~\bibnamefont {Chen}},\
  }\bibfield  {title} {\bibinfo {title} {On the topological surface states of
  the intrinsic magnetic topological insulator {Mn-Bi-Te} family},\ }\href
  {https://doi.org/10.1093/nsr/nwad066} {\bibfield  {journal} {\bibinfo
  {journal} {Natl. Sci. Rev.}\ }\textbf {\bibinfo {volume} {11}},\ \bibinfo
  {pages} {nwad066} (\bibinfo {year} {2023})}\BibitemShut {NoStop}%
\bibitem [{\citenamefont {Hu}\ \emph {et~al.}(2023)\citenamefont {Hu},
  \citenamefont {Qian},\ and\ \citenamefont {Ni}}]{hu2023recent}%
  \BibitemOpen
  \bibfield  {author} {\bibinfo {author} {\bibfnamefont {C.}~\bibnamefont
  {Hu}}, \bibinfo {author} {\bibfnamefont {T.}~\bibnamefont {Qian}},\ and\
  \bibinfo {author} {\bibfnamefont {N.}~\bibnamefont {Ni}},\ }\bibfield
  {title} {\bibinfo {title} {Recent progress in {MnBi$_{2n}$Te$_{3n+1}$}
  intrinsic magnetic topological insulators: crystal growth, magnetism and
  chemical disorder},\ }\href {https://doi.org/10.1093/nsr/nwad282} {\bibfield
  {journal} {\bibinfo  {journal} {Natl. Sci. Rev.}\ }\textbf {\bibinfo {volume}
  {11}},\ \bibinfo {pages} {nwad282} (\bibinfo {year} {2023})}\BibitemShut
  {NoStop}%
\bibitem [{\citenamefont {Sekine}\ and\ \citenamefont
  {Nomura}(2021)}]{sekine2021axion}%
  \BibitemOpen
  \bibfield  {author} {\bibinfo {author} {\bibfnamefont {A.}~\bibnamefont
  {Sekine}}\ and\ \bibinfo {author} {\bibfnamefont {K.}~\bibnamefont
  {Nomura}},\ }\bibfield  {title} {\bibinfo {title} {Axion electrodynamics in
  topological materials},\ }\href {https://doi.org/10.1063/5.0038804}
  {\bibfield  {journal} {\bibinfo  {journal} {J. Appl. Phys.}\ }\textbf
  {\bibinfo {volume} {129}},\ \bibinfo {pages} {141101} (\bibinfo {year}
  {2021})}\BibitemShut {NoStop}%
\bibitem [{\citenamefont {Zhao}\ and\ \citenamefont
  {Liu}(2021)}]{zhao2021routes}%
  \BibitemOpen
  \bibfield  {author} {\bibinfo {author} {\bibfnamefont {Y.}~\bibnamefont
  {Zhao}}\ and\ \bibinfo {author} {\bibfnamefont {Q.}~\bibnamefont {Liu}},\
  }\bibfield  {title} {\bibinfo {title} {Routes to realize the axion-insulator
  phase in {MnBi$_2$Te$_4$(Bi$_2$Te$_3$)$_n$} family},\ }\href
  {https://doi.org/10.1063/5.0059447} {\bibfield  {journal} {\bibinfo
  {journal} {Appl. Phys. Lett.}\ }\textbf {\bibinfo {volume} {119}},\ \bibinfo
  {pages} {060502} (\bibinfo {year} {2021})}\BibitemShut {NoStop}%
\bibitem [{\citenamefont {Gao}\ \emph {et~al.}(2021)\citenamefont {Gao},
  \citenamefont {Liu}, \citenamefont {Hu}, \citenamefont {Qiu}, \citenamefont
  {Tzschaschel}, \citenamefont {Ghosh}, \citenamefont {Ho}, \citenamefont
  {Bérubé}, \citenamefont {Chen}, \citenamefont {Sun}, \citenamefont {Zhang},
  \citenamefont {Zhang}, \citenamefont {Wang}, \citenamefont {Wang},
  \citenamefont {Huang}, \citenamefont {Felser}, \citenamefont {Agarwal},
  \citenamefont {Ding}, \citenamefont {Tien}, \citenamefont {Akey},
  \citenamefont {Gardener}, \citenamefont {Singh}, \citenamefont {Watanabe},
  \citenamefont {Taniguchi}, \citenamefont {Burch}, \citenamefont {Bell},
  \citenamefont {Zhou}, \citenamefont {Gao}, \citenamefont {Lu}, \citenamefont
  {Bansil}, \citenamefont {Lin}, \citenamefont {Chang}, \citenamefont {Fu},
  \citenamefont {Ma}, \citenamefont {Ni},\ and\ \citenamefont
  {Xu}}]{gao2021layer}%
  \BibitemOpen
  \bibfield  {author} {\bibinfo {author} {\bibfnamefont {A.}~\bibnamefont
  {Gao}}, \bibinfo {author} {\bibfnamefont {Y.-F.}\ \bibnamefont {Liu}},
  \bibinfo {author} {\bibfnamefont {C.}~\bibnamefont {Hu}}, \bibinfo {author}
  {\bibfnamefont {J.-X.}\ \bibnamefont {Qiu}}, \bibinfo {author} {\bibfnamefont
  {C.}~\bibnamefont {Tzschaschel}}, \bibinfo {author} {\bibfnamefont
  {B.}~\bibnamefont {Ghosh}}, \bibinfo {author} {\bibfnamefont {S.-C.}\
  \bibnamefont {Ho}}, \bibinfo {author} {\bibfnamefont {D.}~\bibnamefont
  {Bérubé}}, \bibinfo {author} {\bibfnamefont {R.}~\bibnamefont {Chen}},
  \bibinfo {author} {\bibfnamefont {H.}~\bibnamefont {Sun}}, \bibinfo {author}
  {\bibfnamefont {Z.}~\bibnamefont {Zhang}}, \bibinfo {author} {\bibfnamefont
  {X.-Y.}\ \bibnamefont {Zhang}}, \bibinfo {author} {\bibfnamefont {Y.-X.}\
  \bibnamefont {Wang}}, \bibinfo {author} {\bibfnamefont {N.}~\bibnamefont
  {Wang}}, \bibinfo {author} {\bibfnamefont {Z.}~\bibnamefont {Huang}},
  \bibinfo {author} {\bibfnamefont {C.}~\bibnamefont {Felser}}, \bibinfo
  {author} {\bibfnamefont {A.}~\bibnamefont {Agarwal}}, \bibinfo {author}
  {\bibfnamefont {T.}~\bibnamefont {Ding}}, \bibinfo {author} {\bibfnamefont
  {H.-J.}\ \bibnamefont {Tien}}, \bibinfo {author} {\bibfnamefont
  {A.}~\bibnamefont {Akey}}, \bibinfo {author} {\bibfnamefont {J.}~\bibnamefont
  {Gardener}}, \bibinfo {author} {\bibfnamefont {B.}~\bibnamefont {Singh}},
  \bibinfo {author} {\bibfnamefont {K.}~\bibnamefont {Watanabe}}, \bibinfo
  {author} {\bibfnamefont {T.}~\bibnamefont {Taniguchi}}, \bibinfo {author}
  {\bibfnamefont {K.~S.}\ \bibnamefont {Burch}}, \bibinfo {author}
  {\bibfnamefont {D.~C.}\ \bibnamefont {Bell}}, \bibinfo {author}
  {\bibfnamefont {B.~B.}\ \bibnamefont {Zhou}}, \bibinfo {author}
  {\bibfnamefont {W.}~\bibnamefont {Gao}}, \bibinfo {author} {\bibfnamefont
  {H.-Z.}\ \bibnamefont {Lu}}, \bibinfo {author} {\bibfnamefont
  {A.}~\bibnamefont {Bansil}}, \bibinfo {author} {\bibfnamefont
  {H.}~\bibnamefont {Lin}}, \bibinfo {author} {\bibfnamefont {T.-R.}\
  \bibnamefont {Chang}}, \bibinfo {author} {\bibfnamefont {L.}~\bibnamefont
  {Fu}}, \bibinfo {author} {\bibfnamefont {Q.}~\bibnamefont {Ma}}, \bibinfo
  {author} {\bibfnamefont {N.}~\bibnamefont {Ni}},\ and\ \bibinfo {author}
  {\bibfnamefont {S.-Y.}\ \bibnamefont {Xu}},\ }\bibfield  {title} {\bibinfo
  {title} {Layer {Hall} effect in a {2D} topological axion antiferromagnet},\
  }\href {https://doi.org/10.1038/s41586-021-03679-w} {\bibfield  {journal}
  {\bibinfo  {journal} {Nature}\ }\textbf {\bibinfo {volume} {595}},\ \bibinfo
  {pages} {521} (\bibinfo {year} {2021})}\BibitemShut {NoStop}%
\bibitem [{\citenamefont {Chen}\ \emph
  {et~al.}(2022{\natexlab{a}})\citenamefont {Chen}, \citenamefont {Sun},
  \citenamefont {Gu}, \citenamefont {Hua}, \citenamefont {Liu}, \citenamefont
  {Lu},\ and\ \citenamefont {Xie}}]{chen2022layer}%
  \BibitemOpen
  \bibfield  {author} {\bibinfo {author} {\bibfnamefont {R.}~\bibnamefont
  {Chen}}, \bibinfo {author} {\bibfnamefont {H.-P.}\ \bibnamefont {Sun}},
  \bibinfo {author} {\bibfnamefont {M.}~\bibnamefont {Gu}}, \bibinfo {author}
  {\bibfnamefont {C.-B.}\ \bibnamefont {Hua}}, \bibinfo {author} {\bibfnamefont
  {Q.}~\bibnamefont {Liu}}, \bibinfo {author} {\bibfnamefont {H.-Z.}\
  \bibnamefont {Lu}},\ and\ \bibinfo {author} {\bibfnamefont {X.~C.}\
  \bibnamefont {Xie}},\ }\bibfield  {title} {\bibinfo {title} {Layer {Hall}
  effect induced by hidden berry curvature in antiferromagnetic insulators},\
  }\href {https://doi.org/10.1093/nsr/nwac140} {\bibfield  {journal} {\bibinfo
  {journal} {Natl. Sci. Rev.}\ }\textbf {\bibinfo {volume} {11}},\ \bibinfo
  {pages} {nwac140} (\bibinfo {year} {2022}{\natexlab{a}})}\BibitemShut
  {NoStop}%
\bibitem [{\citenamefont {Deng}\ \emph {et~al.}(2020)\citenamefont {Deng},
  \citenamefont {Yu}, \citenamefont {Shi}, \citenamefont {Guo}, \citenamefont
  {Xu}, \citenamefont {Wang}, \citenamefont {Chen},\ and\ \citenamefont
  {Zhang}}]{deng2020quantum}%
  \BibitemOpen
  \bibfield  {author} {\bibinfo {author} {\bibfnamefont {Y.}~\bibnamefont
  {Deng}}, \bibinfo {author} {\bibfnamefont {Y.}~\bibnamefont {Yu}}, \bibinfo
  {author} {\bibfnamefont {M.~Z.}\ \bibnamefont {Shi}}, \bibinfo {author}
  {\bibfnamefont {Z.}~\bibnamefont {Guo}}, \bibinfo {author} {\bibfnamefont
  {Z.}~\bibnamefont {Xu}}, \bibinfo {author} {\bibfnamefont {J.}~\bibnamefont
  {Wang}}, \bibinfo {author} {\bibfnamefont {X.~H.}\ \bibnamefont {Chen}},\
  and\ \bibinfo {author} {\bibfnamefont {Y.}~\bibnamefont {Zhang}},\ }\bibfield
   {title} {\bibinfo {title} {Quantum anomalous hall effect in intrinsic
  magnetic topological insulator {MnBi$_2$Te$_4$}},\ }\href@noop {} {\bibfield
  {journal} {\bibinfo  {journal} {Science}\ }\textbf {\bibinfo {volume}
  {367}},\ \bibinfo {pages} {895} (\bibinfo {year} {2020})}\BibitemShut
  {NoStop}%
\bibitem [{\citenamefont {Liu}\ \emph {et~al.}(2020)\citenamefont {Liu},
  \citenamefont {Wang}, \citenamefont {Li}, \citenamefont {Wu}, \citenamefont
  {Li}, \citenamefont {Li}, \citenamefont {He}, \citenamefont {Xu},
  \citenamefont {Zhang},\ and\ \citenamefont {Wang}}]{liu2020robust}%
  \BibitemOpen
  \bibfield  {author} {\bibinfo {author} {\bibfnamefont {C.}~\bibnamefont
  {Liu}}, \bibinfo {author} {\bibfnamefont {Y.}~\bibnamefont {Wang}}, \bibinfo
  {author} {\bibfnamefont {H.}~\bibnamefont {Li}}, \bibinfo {author}
  {\bibfnamefont {Y.}~\bibnamefont {Wu}}, \bibinfo {author} {\bibfnamefont
  {Y.}~\bibnamefont {Li}}, \bibinfo {author} {\bibfnamefont {J.}~\bibnamefont
  {Li}}, \bibinfo {author} {\bibfnamefont {K.}~\bibnamefont {He}}, \bibinfo
  {author} {\bibfnamefont {Y.}~\bibnamefont {Xu}}, \bibinfo {author}
  {\bibfnamefont {J.}~\bibnamefont {Zhang}},\ and\ \bibinfo {author}
  {\bibfnamefont {Y.}~\bibnamefont {Wang}},\ }\bibfield  {title} {\bibinfo
  {title} {Robust axion insulator and {Chern} insulator phases in a
  two-dimensional antiferromagnetic topological insulator},\ }\href
  {https://doi.org/10.1038/s41563-019-0573-3} {\bibfield  {journal} {\bibinfo
  {journal} {Nat. Mater.}\ }\textbf {\bibinfo {volume} {19}},\ \bibinfo {pages}
  {522} (\bibinfo {year} {2020})}\BibitemShut {NoStop}%
\bibitem [{\citenamefont {Ren}\ \emph {et~al.}(2022)\citenamefont {Ren},
  \citenamefont {Ke}, \citenamefont {Lou},\ and\ \citenamefont
  {Chang}}]{ren2022quantum}%
  \BibitemOpen
  \bibfield  {author} {\bibinfo {author} {\bibfnamefont {Y.}~\bibnamefont
  {Ren}}, \bibinfo {author} {\bibfnamefont {S.}~\bibnamefont {Ke}}, \bibinfo
  {author} {\bibfnamefont {W.-K.}\ \bibnamefont {Lou}},\ and\ \bibinfo {author}
  {\bibfnamefont {K.}~\bibnamefont {Chang}},\ }\bibfield  {title} {\bibinfo
  {title} {Quantum phase transitions driven by sliding in bilayer
  {MnBi$_2$Te$_4$}},\ }\href {https://doi.org/10.1103/PhysRevB.106.235302}
  {\bibfield  {journal} {\bibinfo  {journal} {Phys. Rev. B}\ }\textbf {\bibinfo
  {volume} {106}},\ \bibinfo {pages} {235302} (\bibinfo {year}
  {2022})}\BibitemShut {NoStop}%
\bibitem [{\citenamefont {Zhu}\ \emph {et~al.}(2022)\citenamefont {Zhu},
  \citenamefont {Song}, \citenamefont {Bai}, \citenamefont {Liao},\ and\
  \citenamefont {Pan}}]{zhu2022high}%
  \BibitemOpen
  \bibfield  {author} {\bibinfo {author} {\bibfnamefont {W.}~\bibnamefont
  {Zhu}}, \bibinfo {author} {\bibfnamefont {C.}~\bibnamefont {Song}}, \bibinfo
  {author} {\bibfnamefont {H.}~\bibnamefont {Bai}}, \bibinfo {author}
  {\bibfnamefont {L.}~\bibnamefont {Liao}},\ and\ \bibinfo {author}
  {\bibfnamefont {F.}~\bibnamefont {Pan}},\ }\bibfield  {title} {\bibinfo
  {title} {High chern number quantum anomalous {Hall} effect tunable by
  stacking order in van der {Waals} topological insulators},\ }\href
  {https://doi.org/10.1103/PhysRevB.105.155122} {\bibfield  {journal} {\bibinfo
   {journal} {Phys. Rev. B}\ }\textbf {\bibinfo {volume} {105}},\ \bibinfo
  {pages} {155122} (\bibinfo {year} {2022})}\BibitemShut {NoStop}%
\bibitem [{\citenamefont {Cao}\ \emph {et~al.}(2023)\citenamefont {Cao},
  \citenamefont {Shao}, \citenamefont {Huang}, \citenamefont {Gurung},\ and\
  \citenamefont {Tsymbal}}]{cao2023switchable}%
  \BibitemOpen
  \bibfield  {author} {\bibinfo {author} {\bibfnamefont {T.}~\bibnamefont
  {Cao}}, \bibinfo {author} {\bibfnamefont {D.-F.}\ \bibnamefont {Shao}},
  \bibinfo {author} {\bibfnamefont {K.}~\bibnamefont {Huang}}, \bibinfo
  {author} {\bibfnamefont {G.}~\bibnamefont {Gurung}},\ and\ \bibinfo {author}
  {\bibfnamefont {E.~Y.}\ \bibnamefont {Tsymbal}},\ }\bibfield  {title}
  {\bibinfo {title} {Switchable anomalous hall effects in polar-stacked {2D}
  antiferromagnet {MnBi$_2$Te$_4$}},\ }\href@noop {} {\bibfield  {journal}
  {\bibinfo  {journal} {Nano Letters}\ }\textbf {\bibinfo {volume} {23}},\
  \bibinfo {pages} {3781} (\bibinfo {year} {2023})}\BibitemShut {NoStop}%
\bibitem [{\citenamefont {Ahn}\ \emph {et~al.}(2023)\citenamefont {Ahn},
  \citenamefont {Kang}, \citenamefont {Yoon}, \citenamefont {Ganesh},\ and\
  \citenamefont {Krogel}}]{ahn2023stacking}%
  \BibitemOpen
  \bibfield  {author} {\bibinfo {author} {\bibfnamefont {J.}~\bibnamefont
  {Ahn}}, \bibinfo {author} {\bibfnamefont {S.-H.}\ \bibnamefont {Kang}},
  \bibinfo {author} {\bibfnamefont {M.}~\bibnamefont {Yoon}}, \bibinfo {author}
  {\bibfnamefont {P.}~\bibnamefont {Ganesh}},\ and\ \bibinfo {author}
  {\bibfnamefont {J.~T.}\ \bibnamefont {Krogel}},\ }\bibfield  {title}
  {\bibinfo {title} {Stacking faults and topological properties in
  {MnBi$_2$Te$_4$}: Reconciling gapped and gapless states},\ }\href
  {https://doi.org/10.1021/acs.jpclett.3c01939} {\bibfield  {journal} {\bibinfo
   {journal} {J. Phys. Chem. Lett.}\ }\textbf {\bibinfo {volume} {14}},\
  \bibinfo {pages} {9052} (\bibinfo {year} {2023})}\BibitemShut {NoStop}%
\bibitem [{\citenamefont {Kou}\ \emph {et~al.}(2018)\citenamefont {Kou},
  \citenamefont {Niu}, \citenamefont {Fu}, \citenamefont {Ma}, \citenamefont
  {Yan},\ and\ \citenamefont {Chen}}]{kou2018tunable}%
  \BibitemOpen
  \bibfield  {author} {\bibinfo {author} {\bibfnamefont {L.}~\bibnamefont
  {Kou}}, \bibinfo {author} {\bibfnamefont {C.}~\bibnamefont {Niu}}, \bibinfo
  {author} {\bibfnamefont {H.}~\bibnamefont {Fu}}, \bibinfo {author}
  {\bibfnamefont {Y.}~\bibnamefont {Ma}}, \bibinfo {author} {\bibfnamefont
  {B.}~\bibnamefont {Yan}},\ and\ \bibinfo {author} {\bibfnamefont
  {C.}~\bibnamefont {Chen}},\ }\bibfield  {title} {\bibinfo {title} {{Tunable
  quantum order in bilayer {Bi$_2$Te$_3$}: Stacking dependent quantum spin
  {Hall} states}},\ }\href {https://doi.org/10.1063/1.5038079} {\bibfield
  {journal} {\bibinfo  {journal} {Appl. Phys. Lett.}\ }\textbf {\bibinfo
  {volume} {112}},\ \bibinfo {pages} {243103} (\bibinfo {year}
  {2018})}\BibitemShut {NoStop}%
\bibitem [{\citenamefont {Peng}\ \emph {et~al.}(2020)\citenamefont {Peng},
  \citenamefont {Ma}, \citenamefont {Wang}, \citenamefont {Huang},\ and\
  \citenamefont {Dai}}]{peng2020stacking}%
  \BibitemOpen
  \bibfield  {author} {\bibinfo {author} {\bibfnamefont {R.}~\bibnamefont
  {Peng}}, \bibinfo {author} {\bibfnamefont {Y.}~\bibnamefont {Ma}}, \bibinfo
  {author} {\bibfnamefont {H.}~\bibnamefont {Wang}}, \bibinfo {author}
  {\bibfnamefont {B.}~\bibnamefont {Huang}},\ and\ \bibinfo {author}
  {\bibfnamefont {Y.}~\bibnamefont {Dai}},\ }\bibfield  {title} {\bibinfo
  {title} {Stacking-dependent topological phase in bilayer
  $m\mathrm{B}{\mathrm{i}}_{2}\mathrm{T}{\mathrm{e}}_{4}(m=\mathrm{Ge},\mathrm{Sn},\mathrm{Pb})$},\
  }\href {https://doi.org/10.1103/PhysRevB.101.115427} {\bibfield  {journal}
  {\bibinfo  {journal} {Phys. Rev. B}\ }\textbf {\bibinfo {volume} {101}},\
  \bibinfo {pages} {115427} (\bibinfo {year} {2020})}\BibitemShut {NoStop}%
\bibitem [{SM()}]{SM}%
  \BibitemOpen
  \href@noop {} {}\bibinfo {note} {See Supplemental Material for detailed
  calculation methods, electronic band structures of ABC- and CAC-stacking MBT
  thin films with 3-8 SLs, and tight-binding parameters of
  Hamiltonians.}\BibitemShut {Stop}%
\bibitem [{\citenamefont {Lian}\ \emph {et~al.}(2020)\citenamefont {Lian},
  \citenamefont {Liu}, \citenamefont {Zhang},\ and\ \citenamefont
  {Wang}}]{lian2020flat}%
  \BibitemOpen
  \bibfield  {author} {\bibinfo {author} {\bibfnamefont {B.}~\bibnamefont
  {Lian}}, \bibinfo {author} {\bibfnamefont {Z.}~\bibnamefont {Liu}}, \bibinfo
  {author} {\bibfnamefont {Y.}~\bibnamefont {Zhang}},\ and\ \bibinfo {author}
  {\bibfnamefont {J.}~\bibnamefont {Wang}},\ }\bibfield  {title} {\bibinfo
  {title} {Flat chern band from twisted bilayer
  {${\mathrm{MnBi}}_{2}{\mathrm{Te}}_{4}$}},\ }\href
  {https://doi.org/10.1103/PhysRevLett.124.126402} {\bibfield  {journal}
  {\bibinfo  {journal} {Phys. Rev. Lett.}\ }\textbf {\bibinfo {volume} {124}},\
  \bibinfo {pages} {126402} (\bibinfo {year} {2020})}\BibitemShut {NoStop}%
\bibitem [{\citenamefont {Zhang}\ \emph {et~al.}(2009)\citenamefont {Zhang},
  \citenamefont {Liu}, \citenamefont {Qi}, \citenamefont {Dai}, \citenamefont
  {Fang},\ and\ \citenamefont {Zhang}}]{zhang2009topological}%
  \BibitemOpen
  \bibfield  {author} {\bibinfo {author} {\bibfnamefont {H.}~\bibnamefont
  {Zhang}}, \bibinfo {author} {\bibfnamefont {C.-X.}\ \bibnamefont {Liu}},
  \bibinfo {author} {\bibfnamefont {X.-L.}\ \bibnamefont {Qi}}, \bibinfo
  {author} {\bibfnamefont {X.}~\bibnamefont {Dai}}, \bibinfo {author}
  {\bibfnamefont {Z.}~\bibnamefont {Fang}},\ and\ \bibinfo {author}
  {\bibfnamefont {S.-C.}\ \bibnamefont {Zhang}},\ }\bibfield  {title} {\bibinfo
  {title} {Topological insulators in {Bi$_2$Se$_3$, Bi$_2$Te$_3$ and
  Sb$_2$Te$_3$} with a single {Dirac} cone on the surface},\ }\href
  {https://doi.org/10.1038/nphys1270} {\bibfield  {journal} {\bibinfo
  {journal} {Nature Physics}\ }\textbf {\bibinfo {volume} {5}},\ \bibinfo
  {pages} {438} (\bibinfo {year} {2009})}\BibitemShut {NoStop}%
\bibitem [{\citenamefont {Liu}\ \emph {et~al.}(2010)\citenamefont {Liu},
  \citenamefont {Qi}, \citenamefont {Zhang}, \citenamefont {Dai}, \citenamefont
  {Fang},\ and\ \citenamefont {Zhang}}]{liu2010model}%
  \BibitemOpen
  \bibfield  {author} {\bibinfo {author} {\bibfnamefont {C.-X.}\ \bibnamefont
  {Liu}}, \bibinfo {author} {\bibfnamefont {X.-L.}\ \bibnamefont {Qi}},
  \bibinfo {author} {\bibfnamefont {H.}~\bibnamefont {Zhang}}, \bibinfo
  {author} {\bibfnamefont {X.}~\bibnamefont {Dai}}, \bibinfo {author}
  {\bibfnamefont {Z.}~\bibnamefont {Fang}},\ and\ \bibinfo {author}
  {\bibfnamefont {S.-C.}\ \bibnamefont {Zhang}},\ }\bibfield  {title} {\bibinfo
  {title} {Model hamiltonian for topological insulators},\ }\href
  {https://doi.org/10.1103/PhysRevB.82.045122} {\bibfield  {journal} {\bibinfo
  {journal} {Phys. Rev. B}\ }\textbf {\bibinfo {volume} {82}},\ \bibinfo
  {pages} {045122} (\bibinfo {year} {2010})}\BibitemShut {NoStop}%
\bibitem [{\citenamefont {Chen}\ \emph
  {et~al.}(2022{\natexlab{b}})\citenamefont {Chen}, \citenamefont {Ye},
  \citenamefont {Zou}, \citenamefont {Gu}, \citenamefont {Xu},\ and\
  \citenamefont {Duan}}]{chen2022basic}%
  \BibitemOpen
  \bibfield  {author} {\bibinfo {author} {\bibfnamefont {H.}~\bibnamefont
  {Chen}}, \bibinfo {author} {\bibfnamefont {M.}~\bibnamefont {Ye}}, \bibinfo
  {author} {\bibfnamefont {N.}~\bibnamefont {Zou}}, \bibinfo {author}
  {\bibfnamefont {B.-L.}\ \bibnamefont {Gu}}, \bibinfo {author} {\bibfnamefont
  {Y.}~\bibnamefont {Xu}},\ and\ \bibinfo {author} {\bibfnamefont
  {W.}~\bibnamefont {Duan}},\ }\bibfield  {title} {\bibinfo {title} {Basic
  formulation and first-principles implementation of nonlinear magneto-optical
  effects},\ }\href {https://doi.org/10.1103/PhysRevB.105.075123} {\bibfield
  {journal} {\bibinfo  {journal} {Phys. Rev. B}\ }\textbf {\bibinfo {volume}
  {105}},\ \bibinfo {pages} {075123} (\bibinfo {year}
  {2022}{\natexlab{b}})}\BibitemShut {NoStop}%
\bibitem [{\citenamefont {Sivadas}\ \emph {et~al.}(2016)\citenamefont
  {Sivadas}, \citenamefont {Okamoto},\ and\ \citenamefont
  {Xiao}}]{sivadas2016gate}%
  \BibitemOpen
  \bibfield  {author} {\bibinfo {author} {\bibfnamefont {N.}~\bibnamefont
  {Sivadas}}, \bibinfo {author} {\bibfnamefont {S.}~\bibnamefont {Okamoto}},\
  and\ \bibinfo {author} {\bibfnamefont {D.}~\bibnamefont {Xiao}},\ }\bibfield
  {title} {\bibinfo {title} {Gate-controllable magneto-optic kerr effect in
  layered collinear antiferromagnets},\ }\href
  {https://doi.org/10.1103/PhysRevLett.117.267203} {\bibfield  {journal}
  {\bibinfo  {journal} {Phys. Rev. Lett.}\ }\textbf {\bibinfo {volume} {117}},\
  \bibinfo {pages} {267203} (\bibinfo {year} {2016})}\BibitemShut {NoStop}%
\end{thebibliography}%
\end{document}